\documentclass[aps,prx,twocolumn,citeautoscript,nofootinbib]{revtex4-2}
\usepackage{silence}
\usepackage{amsmath}

\usepackage{feynmp}
\usepackage{comment}
\usepackage{subcaption}

\usepackage{amssymb}
\usepackage{bbm}
\usepackage{braket}
\usepackage{xcolor}
\usepackage{pifont}
\usepackage{slashed}
\usepackage[mathscr]{euscript}
\usepackage[shortlabels]{enumitem}
\usepackage{tikz-feynman}
\usepackage[papersize={8.5in,11in}]{geometry}
\allowdisplaybreaks
\usepackage{slashed}
\usepackage{simpler-wick}
\usepackage{framed}
\usepackage{graphicx}
\usepackage{subcaption}
\usepackage[colorlinks=true]{hyperref}
\usepackage{mathtools}
\renewcommand{\approx}{\simeq}

\renewcommand{\Im}{\text{Im}}

\newcommand{\ce}{\mathcal{E}}

\definecolor{wrongultramarine}{rgb}{1,0.5,0}

\newcommand{\rd}{{\rm d}}

\newcommand{\Tr}{{\rm \, Tr\,}}
\newcommand{\beq}{\begin{equation}}
\newcommand{\eeq}{\end{equation}}
\def\bea{\begin{eqnarray}}
\def\eea{\end{eqnarray}}

\newcommand{\calG}{\mathcal{G}}

\hypersetup{
    bookmarks=true,         
    unicode=false,          
    pdftoolbar=true,        
    pdfmenubar=true,        
    pdffitwindow=false,     
    pdfstartview={FitH},    
    pdfsubject={},   
    pdfcreator={},   
    pdfproducer={}, 
    pdfkeywords={} {} {}, 
    pdfnewwindow=true,      
    colorlinks=true,       
    linkcolor=blue, 
    citecolor=blue,        
    filecolor=magenta,      
    urlcolor=blue           
}

\tikzset{
  mid arrow/.style={postaction={decorate,decoration={
        markings,
        mark=at position .575 with {\arrow[#1]{stealth}}
      }}},
  near arrow/.style={postaction={decorate,decoration={
        markings,
        mark=at position .275 with {\arrow[#1]{stealth}}
      }}},
   far arrow/.style={postaction={decorate,decoration={
        markings,
        mark=at position .800 with {\arrow[#1]{stealth}}
      }}},
}

\usepackage{pdfpages}
\usepackage{pgffor}
\makeatletter
\AtBeginDocument{\let\LS@rot\@undefined}
\makeatother

\begin{document}
\title{Phonon Thermal Hall Effect in a non-Kramers Paramagnet}

\author{Haoyu Guo}
\affiliation{Laboratory of Atomic and Solid State Physics, Cornell University,
142 Sciences Drive, Ithaca NY 14853-2501, USA}
\affiliation{Department of Physics, Harvard University, Cambridge MA 02138, USA}

\date{\today}

\begin{abstract}
    With an increasing number of experiments observing phonon thermal Hall effect in magnets, it is theoretically desired to have a minimal model for the phonon thermal Hall effect. In this work, we study a simple model of acoustic phonons and uncorrelated paramagnetic doublets. When the doublets are non-Kramers, a linear pseudospin-phonon coupling is allowed by time-reversal symmetry and yields resonant phonon-pseudospin scattering. We compute the thermal longitudinal and Hall conductivities of the model and found qualitative agreement with recent experiments in $\rm{Pr}_2\rm{Ir}_2\rm{O}_7$ [Uehara {\it et al}.,Nat Commun {\bf13}, 4604 (2022)].
\end{abstract}

\maketitle


\section{Introduction}
Thermal Hall effect refers to the phenomenon that a heat current can be steered by the external magnetic field. Recently, thermal Hall effects of phononic origin have been observed in a wide range materials including cuprates \cite{Grissonnanche2019,Grissonnanche2020,Boulanger2020,Boulanger2022}, magnetic insulator $\rm{Cu}_3\rm{Te}\rm{O}_6$ \cite{Chen2022}, ferroelectric $\rm{Sr}\rm{Ti}\rm{O}_3$ \cite{Li2020}, spin ice \cite{Uehara2022, Hirschberger2015, Hirokane2019}  and spin liquid candidate $\alpha$-$\rm{Ru}\rm{Cl}_3$ \cite{Lefrancois2022}. Being charge neutral, the thermally active acoustic phonon at low temperature cannot directly couple to magnetic field. To acquire chirality, they can only couple to excitations in the system, such as charged defects\cite{Flebus2022a}, dynamical spin defects \cite{Guo2022a}, electric dipole fluctuations \cite{Chen2020}, magnons \cite{Ye2021,Mangeolle2022,Mangeolle2022a} or fractionalized spinons \cite{Zhang2021}. Therefore, the phonon thermal Hall effect has emerged as a new probe for magnetic excitations in the system.

On the theoretical side, an important question to ask is what kind of correlation is probed by the phonon thermal Hall effect. Recently, it is proposed by Mangeolle {\it et al.} \cite{Mangeolle2022,Mangeolle2022a} that the phonon thermal Hall effect detects an out-of-time ordered correlator of the collective excitations. 

In this paper, we answer the above question by studying a minimal model: a two-level paramagnet. We show that this very simple magnetic system, in a non-Kramers system, can be coupled to phonons and produce phonon thermal Hall effect. The operators in the two-level Hilbert space can be represented using pseudospin Pauli matrices. We will show that the phonon thermal Hall effect can probe the usual two-point pseudospin correlation function of pseudospins. In the second half of the manuscript, we will apply our theory to the spin ice system $\rm{Pr}_2\rm{Ir}_2\rm{O}_7$ \cite{Uehara2022} where the phonon thermal Hall effect has recently been measured.

In the experimental regime where phonon thermal Hall effect is observed, phonons can be treated semiclassically as long-lived quasiparticles. The mechanism of electron Hall effect has been studied both in the frameworks of Boltzmann equation and Kubo formalism \cite{Sinitsyn2007,Nagaosa2010} with agreeing results, and can be generalized to other quasiparticles including phonons. Accordingly, phonons can contribute to Hall effect in three mechanisms: intrinsic Berry curvature \cite{Qin2012}, skew scattering \cite{Mori2014,Chen2020,Guo2021,Sun2022,Flebus2022a} and side jump\cite{Guo2022a}. We will argue that for the paramagnet model we consider, only the side-jump mechanism is important if we assume inversion symmetry of the system.


The authors previously developed the side-jump theory of phonon thermal Hall effect in cuprates \cite{Guo2022a}, which involves phonon scattering on dynamical defects inducing phonon-spin coupling in a time-reversal breaking environment. However, this work focuses on non-Kramers doublets where phonon-pseudospin coupling is determined by time-reversal and lattice symmetry. The theory goes beyond the previous work by incorporating feedback from pseudospin to phonons and providing a coherent description of both longitudinal thermal conductivity $\kappa_{xx}$ and thermal Hall conductivity $\kappa_{xy}$ via resonant phonon-pseudospin scattering. Recent experiments in $\rm{Pr}_2\rm{Ir}_2\rm{O}_7$ \cite{Uehara2022} measured a significant thermal Hall effect due to phonons, and our simple model is in good qualitative agreement with the results.

\section{The acoustic phonon} We work in units $\hbar=k_B=1$. 
Following notations in \cite{Guo2022a}, on a lattice, the phonon Hamiltonian can be written as
\begin{equation}\label{eq:Hph}
  H_\text{ph}=\sum_{p,i} \frac{\pi^i_p \pi^i_p}{2m}+\frac{1}{2} \sum_{pqij} u_p^i C_{pq}^{ij}u_q^j\,.
\end{equation} Here $u_p^i$ is the $i$-th component of the lattice displacement at lattice site $p$, and $\pi_p^i$ is its conjugate momentum satisfying $[u_p^i,\pi_q^j]=i\delta_{pq}\delta^{ij}$. 
As a simplification we choose the acoustic phonon to have an isotropic dispersion, meaning the elasticity matrix $C_{pq}^{ij}$, after fourier transform, takes the form
\begin{equation}\label{}
  C^{ij}(k)=m c_T^2 \delta^{ij}k^2+m(c_L^2-c_T^2)k^i k^j\,.
\end{equation} Here $m$ is the ion mass and $c_L(c_T)$ is the velocity of the longitudinal (transverse) phonon.

\section{The pseudospin Hamiltonian}
    We consider a spatially uncorrelated paramagnet, with the Hamiltonian
\begin{equation}\label{}
  H_\text{spin}=-\sum_p \frac{\Delta}{2}\sigma_p^3\,.
\end{equation} Here $\sigma_p^3$ denotes the third Pauli matrix operator residing at lattice site $p$. The gap $\Delta$ is controlled by Zeeman energy, having the form
\begin{equation}\label{}
  \Delta\propto\mu_B {H}\,.
\end{equation} Here $H$ is the physical external magnetic field, and $\mu_B$ denotes Bohr's magneton. 

\section{Time-reversal symmetry and phonon-pseudospin coupling}
Acoustic phonon, as a Goldstone mode, couples to pseudospins at leading order through
\begin{equation}\label{eq:Hspinph_general}
  H_\text{spin-ph}=\sum_{pij\alpha} K_{ij\alpha}\partial_i u^j_p \sigma_p^\alpha\,.
\end{equation}  Here $K_{ij\alpha}$ is the coupling constant and $\alpha$ denotes the components of the pseudospin $\sigma_p$ at lattice site $p$.
To maintain time-reversal symmetry $\mathcal{T}$, only $\sigma^3$ is odd while $\sigma^1$ and $\sigma^2$ are even under $\mathcal{T}$. This feature is satisfied by the non-Kramers doublet with $\mathcal{T}=\sigma^1 C$ where $C$ is complex conjugation, which can arise from an atomic state with even number of unpaired electrons. Microscopically, the coupling \eqref{eq:Hspinph_general} arise directly from Coulomb interaction (Appendix.~\ref{app:spin-ph-cpl}).

The coupling in \eqref{eq:Hspinph_general} violates time-reversal symmetry for Kramers doublet with $\mathcal{T}=i\sigma^2 C$, where $\mathcal{T}^2=-1$. Instead, the leading order coupling is proportional to ion momenta and can be expressed as $H_\text{spin-ph}=\sum_{pi\alpha} K_{i\alpha}\pi^i_p \sigma_p^\alpha$. This originates from the magnetic field felt by the paramagnetic moment from the moving ions, which is much weaker as the ions are heavy and non-relativistic.
From now on, we will consider only non-Kramers paramagnet with coupling \eqref{eq:Hspinph_general}.

\section{Resonant Scattering and Phonon lifetime} Now, we study the transport properties of the model $H=H_\text{ph}+H_\text{spin}+H_\text{spin-ph}$. One consequence of the linear pseudospin-phonon coupling given in Eq.\eqref{eq:Hspinph_general} is the resonant scattering between the phonons and the pseudospins.
The single particle decay rate of the phonon can be extracted from the imaginary part of phonon self energy.
In Appendix.~\ref{app:resscat} we show using Abrikosov fermions \cite{Joshi2020} that the first contribution appears at fourth order in the coupling constants $K_{ij\alpha}$, with the form ($\beta=1/T$ is inverse temperature)
\begin{equation}\label{eq:Gammaph}
\footnotesize
\begin{split}
  &\Gamma_\text{res}(\omega)=R\omega^4\Bigg[\frac{\Delta^2+\omega^2}{(\Delta^2-\omega^2)^2}\left(1-\tanh^2\left(\frac{\beta \Delta}{2}\right)\right)\Bigg]\,,\\
  &R=\frac{K^{(4)}}{(mc^2)^2 \omega_D^3}\,.
\end{split}
\end{equation} Here we have averaged the decay rate over directions of phonon momentum, which is justified assuming phonon-phonon scattering averages lifetime \cite{Ziman2001}.
In Eq.\eqref{eq:Gammaph}, $K^{(4)}$ is a complicated fourth order combination of couplings constants $K_{ij\alpha}$, and is generically nonzero; $c\sim c_L,c_T$ is the typical phonon velocity; $\omega_D$ is Debye frequency. The result we have obtained here is consistent with earlier computations using other diagrammatic techniques and Born's approximation \cite{Toombs1973,Sheard1973,Toombs1985}. The divergent scattering rate in Eq.\eqref{eq:Gammaph} is not an issue because transport coefficients stay finite and are insensitive to the value of $\Gamma_\text{res}$ at resonance \cite{Pohl1962}.

\section{Phonon Thermal Hall effect}
We argue that only the side jump contribution is significant. The Berry curvature contribution is small compared to the other two given that phonon mean-free path is much longer than lattice spacing \cite{Chen2020,Guo2021,Sun2022,Guo2022a}. Next, the skew scattering mechanism is suppressed due to a parity mismatch in skew scattering and the fact that Eq.\eqref{eq:Hspinph_general} is inversion even \cite{Mori2014,Guo2021,Sun2022} (see Appendix.~\ref{app:skew}).

Finally, we consider the side-jump mechanism, which refers to the process that when a particle is scattered by the pseudospin from one band to another band, its coordinate is shifted due to the difference of Berry connection \cite{Sinitsyn2006}. The side-jump effect contributes to the phonon velocity by multiplying the coordinate shift with scattering rate, and can therefore contribute to thermal Hall effect. The side-jump thermal Hall effect can be schematically written as
\begin{equation}\label{}
  \kappa_{xy}^{sj}\sim \frac{1}{3}Cc c^{sj}\tau_\text{ph}\,.
\end{equation} Here $C$ is the heat capacity of the contributing phonon modes, $c$ is the usual sound velocity, $c^{sj}$ is the side-jump velocity and $\tau_{\text{ph}}$ is total mean-free time of the phonons.

In Appendix.~\ref{app:the} we review the side-jump theory of the phonon thermal Hall effect \cite{Guo2022a} and explain how it can be extended to capture the feedback of resonant scattering. The result is
\begin{equation}\label{eq:kxyint}
\begin{split}
  &\kappa_{xy}=\frac{1}{30\pi m}\sum_{a=L,T}\frac{\tanh\left(\frac{\beta \Delta}{2}\right)}{c_a^3}  \\
  &\int_0^\infty \rd \omega(-\beta n_B'(\omega)) \omega^4\frac{\pi^{-1}K_a^{}}{(\omega-\Delta)^2+\frac{1}{4}\Gamma_a(\omega)^2}\,.
\end{split}
\end{equation} Here $a=L,T$ denotes the phonon bands; $K_L$ and $K_T$ are quadratic antisymmetric combinations of coupling constants, given by Eqs. (17) and (18) of \cite{Guo2022a} respectively (also reproduced in Appendix.~\ref{app:the}). The phonon scattering rate $\Gamma_a(\omega)=\Gamma_\text{res}(\omega)+\Gamma_\text{non-res}(\omega)$ is the sum of resonant contribution \eqref{eq:Gammaph} and non-resonant contribution specified later.
The result below assumes $\Gamma_a(\omega)$ is independent of $a$:
\begin{equation}\label{eq:kxyeff}
  \kappa_{xy}=\frac{1}{30\pi m} \frac{\Delta^4\left(c_L^{-3}K_L^{}+c_T^{-3}K_T^{}\right)}{T^2 \sinh\left(\Delta/T\right)}\frac{1}{\Gamma_\text{eff}}\,.
\end{equation} Here
$$\Gamma_\text{eff}^{-1}=\frac{2^{2/3}}{3\left[R \Delta^4 \left(1-\tanh^2\left(\frac{\beta \Delta}{2}\right)\right)\right]^{1/3}}\,,$$ if $\Gamma_\text{res}$ dominates over $\Gamma_\text{non-res}$ ($R\to\infty$), and in the opposite limit $\Gamma_\text{eff}=\Gamma_\text{non-res}(\Delta)$.

\begin{figure}[htb]
  \centering
  \includegraphics[width=\columnwidth]{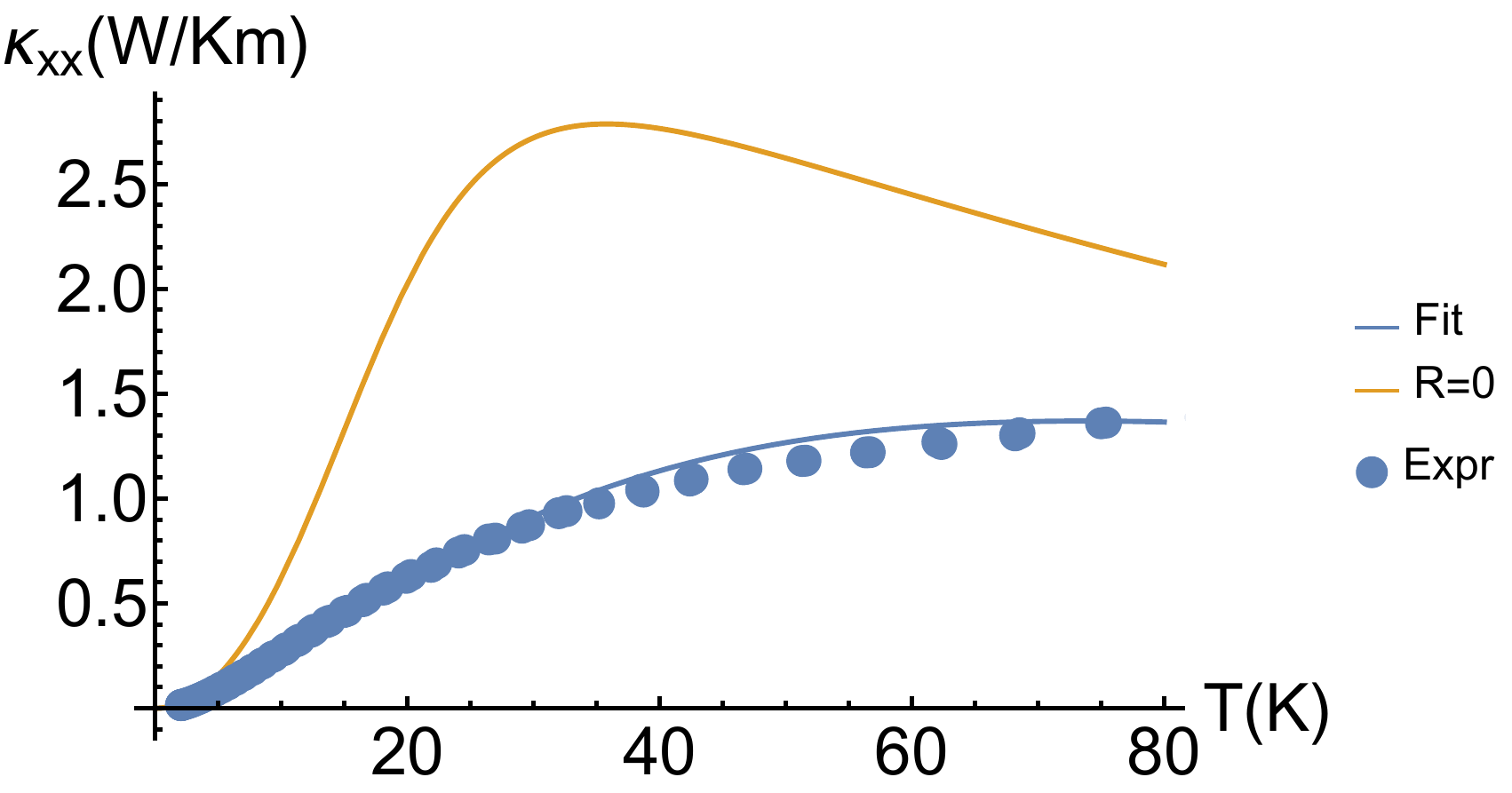}
  \caption{Thermal conductivity $\kappa_{xx}$ at zero field in the $(001)$ plane. The experimental data is adapted from \cite{Uehara2022}. The blue curve is the Debye-Callaway model fit. The yellow curve is the model with pseudospin-phonon coupling removed and other parameters held fixed. Note the reappearance of phonon peak.}\label{fig:kxxT}
\end{figure}

\begin{figure*}[htb]
    \centering
    \includegraphics[width=0.92\textwidth]{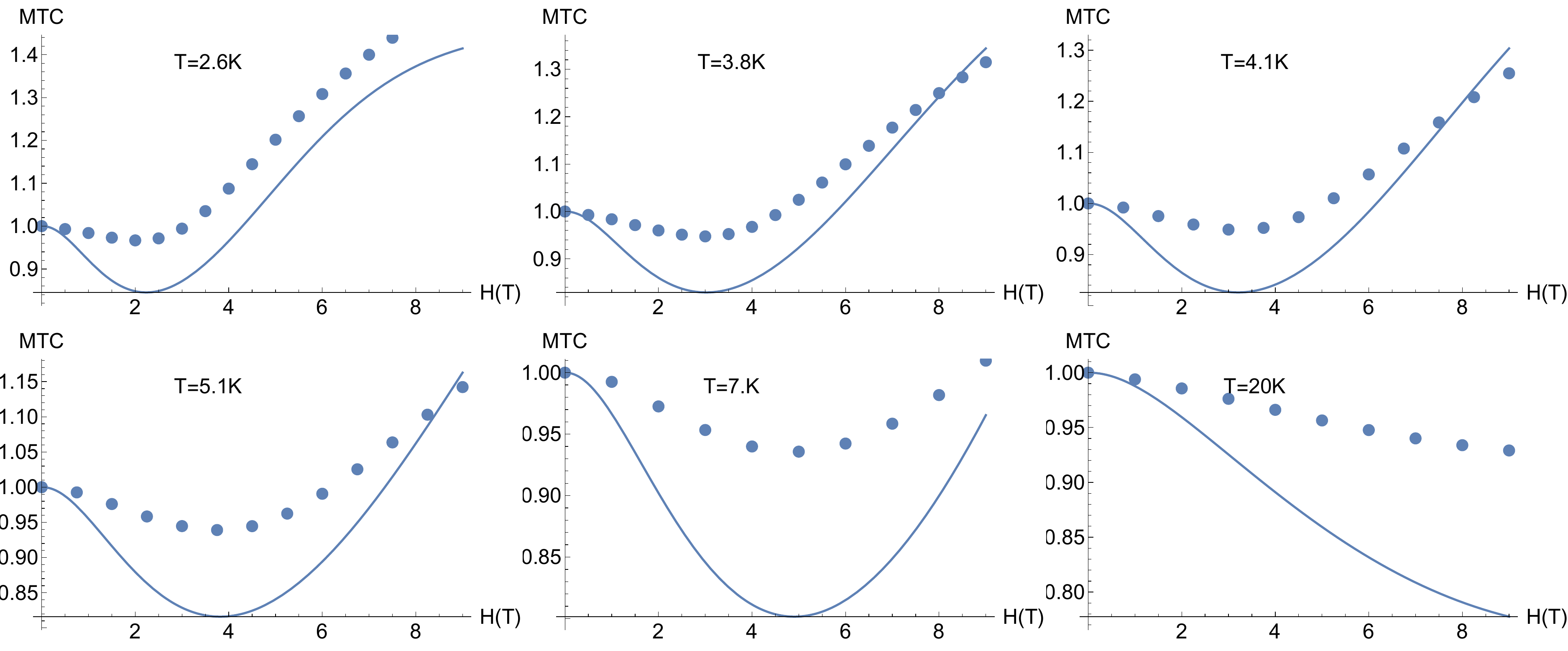}
    \caption{Magneto-thermal conductivity $\text{MTC}=\frac{\kappa_{xx}(H)}{\kappa_{xx}(0)}$ with field $H\parallel[001]$. The data points are adapted from \cite{Uehara2022} and the curves are fits based on Debye-Callaway model. }\label{fig:MTC001}
\end{figure*}

\section{Comparison with experiment}

Our theory is compared to experiments in $\rm{Pr}_2\rm{Ir}_2\rm{O}_7$ \cite{Uehara2022}, a material with a pyrochlore structure and the magnetism is dominated by the moments of the $\rm{Pr}^{3+}$ ions. This ion contains two $4f$ electrons, and its ground state is a non-Kramers doublet with ising anisotropy along the $[111]$ direction, protected by a large gap $(162{\rm K})$ \cite{Machida2005}. The pseudospins are ferromagnetically coupled with an exchange constant of  $2J_\text{eff}=1.4\rm{K}$ \cite{Machida2010,Onoda2011}. As this is smaller than the temperature and Zeeman energy where thermal Hall effect is measured, we can treat the $\rm{Pr}^{3+}$ moments as paramagnetic without considering correlations at zeroth order.

The phonon-pseudospin coupling in $\rm{Pr}_2\rm{Ir}_2\rm{O}_7$ can be refined by taking the local $D_{3d}$ symmetry near $\rm{Pr}^{3+}$ ions into account, with the result \cite{Patri2020}
\begin{equation}\label{eq:Hspin-ph}
  \begin{split}
    H_\text{spin-ph}&=k_1\left[\sigma^1\left(\partial_1 u^1-\partial_2 u^2\right)+\sigma^2(\partial_1u^2+\partial_2 u^1)\right]\\
               +k_{2a}&\left[\sigma^1 \partial_1 u^3-\sigma^2 \partial_2 u^3\right]+k_{2b}\left[\sigma^1 \partial_3 u^1-\sigma^2 \partial_3 u^2\right]\,.
  \end{split}
  \end{equation}Here $k_1,k_{2a}$ and $k_{2b}$ are couplings that cannot be fixed by symmetry. The coordinate here is aligned with the local frame of $\rm{Pr}^{3+}$ ion (Appendix.~\ref{app:frame}). We can also derive it by perturbing the crystal electric field (CEF) by lattice displacement (Appendix.~\ref{app:spin-ph-cpl}), leading to an order-of-magnitude estimate of $10^2$K-$10^3$K.

  The strong phonon-pseudospin coupling \eqref{eq:Hspin-ph} influences both $\kappa_{xx}$ and $\kappa_{xy}$ in pyrochlore compounds. In Kramers systems $\rm{Yb}_2\rm{Ti}_2\rm{O}_7$, $\rm{Dy}_2\rm{Ti}_2\rm{O}_7$ and the non-magnetic $\rm{Y}_2\rm{Ti}_2\rm{O}_7$, pronounced phonon peaks are observed \cite{Tokiwa2016,Kolland2013} in $\kappa_{xx}(T,H=0)$. In contrast, the non-Kramers $\rm{Tb}_2\rm{Ti}_2\rm{O}_7$ and $\rm{Pr}_2\rm{Ir}_2\rm{O}_7$ show no phonon peak \cite{Li2013,Uehara2022} but instead exhibit thermal Hall effect \cite{Hirschberger2015,Hirokane2019,Uehara2022}.

  The thermal transport data in \cite{Uehara2022} supports a paramagnetic pseudospin picture. The magneto-thermal conductivity (MTC) $\kappa_{xx}(H)$ shows suppression $\kappa_{xx}(H)<\kappa_{xx}(0)$ at low field and enhancement $\kappa_{xx}(H)>\kappa_{xx}(0)$ at high field, consistent with Eq.\eqref{eq:Gammaph}. The thermal Hall effect shows a scaling collapse of data $\kappa_{xy}(H,T)=C(T)f(H/T)$, indicating a single energy scale set by the magnetic field $H$, which agrees with Eq.\eqref{eq:kxyeff}. 

\begin{figure*}[htb]
    \centering
    \includegraphics[width=0.92\textwidth]{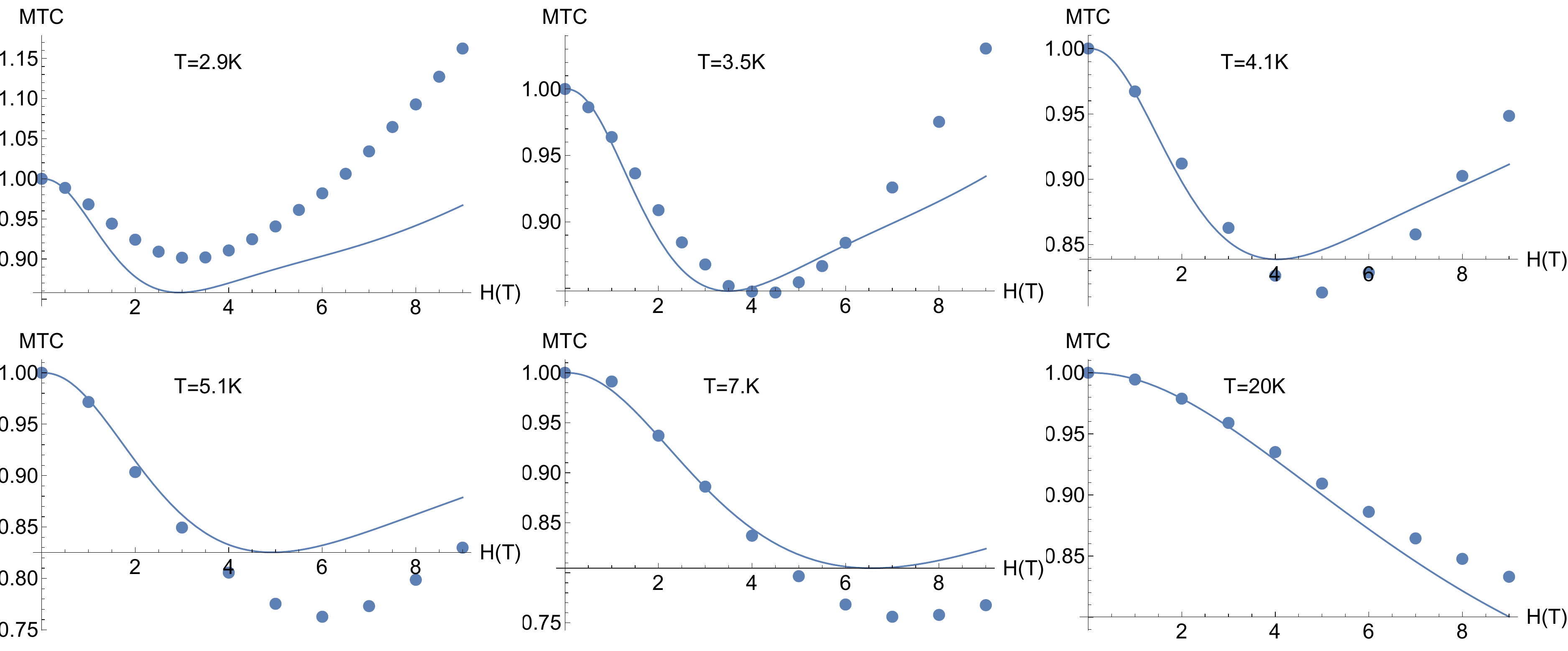}
    \caption{Similar to Fig.~\ref{fig:MTC001} but with field direction $H\parallel [111]$. }\label{fig:MTC111}
\end{figure*}

    To quantitatively compare with the experimental data, we use the Debye-Callaway model \cite{Ziman2001} to fit the longitudinal thermal conductivity
\begin{equation}\label{}
  \kappa_{xx}=\frac{k_B^4}{2\pi^2 c\hbar^3}T^3\int_0^{T_D/T}\rd x \frac{x^4e^x}{(e^x-1)^2\Gamma(xT)}\,,
\end{equation} where we take the phonon scattering rate $\Gamma=\Gamma_\text{res}+\Gamma_\text{non-res}$ with $\Gamma_\text{res}$ given by \eqref{eq:Gammaph}, and
\begin{equation}\label{eq:Gammans}
  \Gamma_\text{non-res}(\omega)=\frac{c}{L_B}+A_0\omega+A_1\omega^4+A_2 \omega^2T\exp\left(-\frac{T_D}{b T}\right)\,,
\end{equation} which describes boundary scattering, line defect scattering, point defect scattering and phonon-phonon Umklapp scattering respectively \cite{Ziman2001}.
The Debye temperature $T_D=436\rm{K}$ is determined from specific heat measurement \cite{Ghosh2021}, and combining with cubic cell lattice constant $a=10.4 \textup{~\AA}$ \cite{Millican2007}, we obtain the speed of sound $c=3424{\rm m/s}$. The boundary scattering mean-free path is fixed at $L_B=1{\rm mm}$ from sample size. The parameter $b$ is fixed empirically \cite{Yang2022} to $b=2N^{1/3}$ where $N=88$ is the number of atoms in a cubic cell.
We identify four fitting parameters: $A_0,A_1,A_2$ and $gJ$. The remaining fitting parameter(s) is the scattering strength $R$ (see \eqref{eq:Gammaph}). We need to assign one $R$ to each inequivalent sublattice of $\rm{Pr}^{3+}$.

\begin{figure}[htb]
  \centering
  \includegraphics[width=\columnwidth]{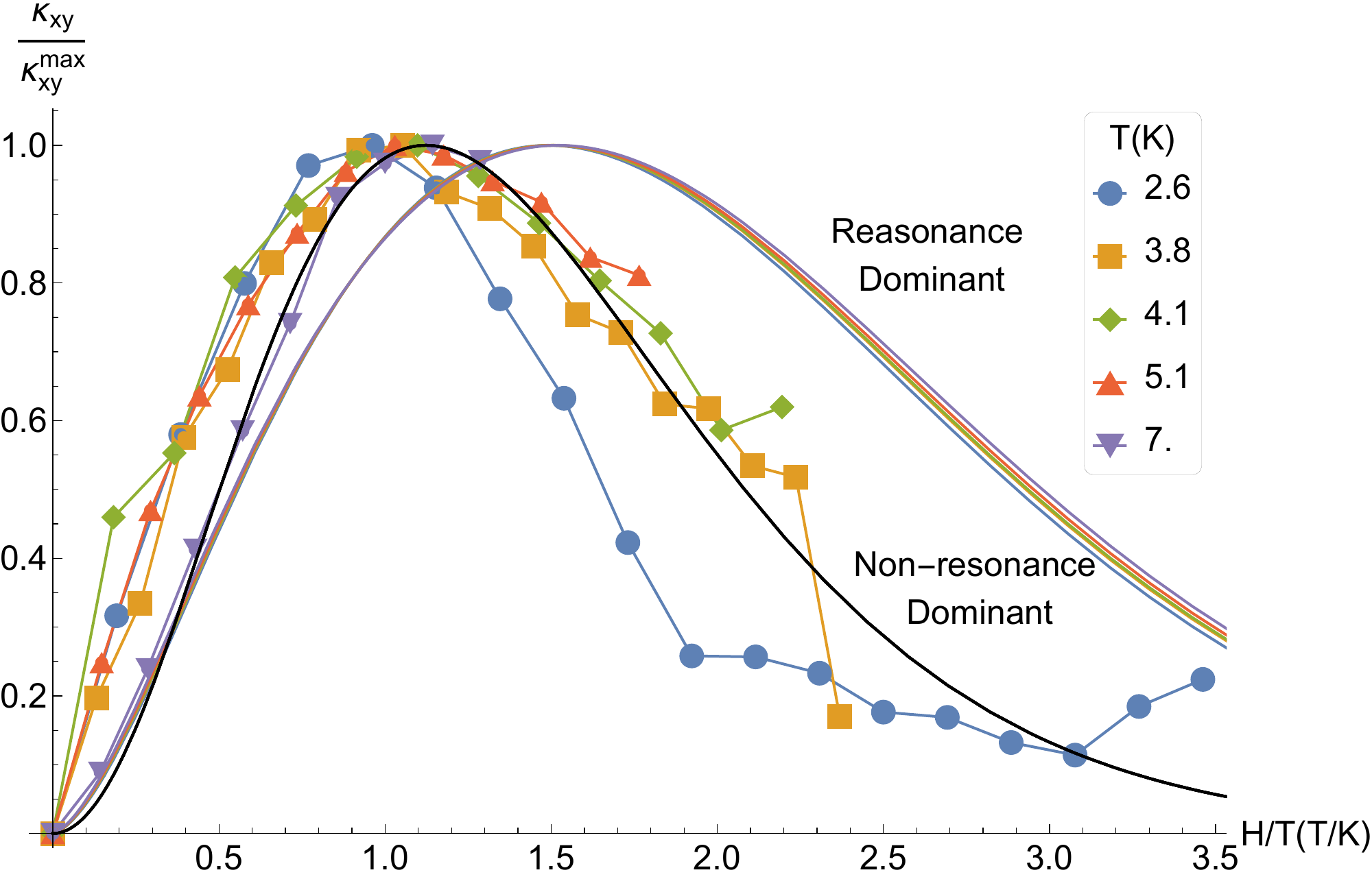}
  \caption{Normalized thermal Hall conductivity when $H\parallel[001]$. The data is from \cite{Uehara2022}. The colored lines are computed using Eq.\eqref{eq:kxyint} with extracted fitting parameters, where resonant scattering dominates $\Gamma_\text{eff}$. The spread of different colors reflect the influence of temperature dependent $\Gamma_\text{eff}$. The black line is the theoretical curve assuming non-resonant scattering is dominant in $\Gamma_\text{eff}$ $(R\to0)$.}\label{fig:kxy001}
\end{figure}
\begin{figure}[htb]
  \centering
  \includegraphics[width=\columnwidth]{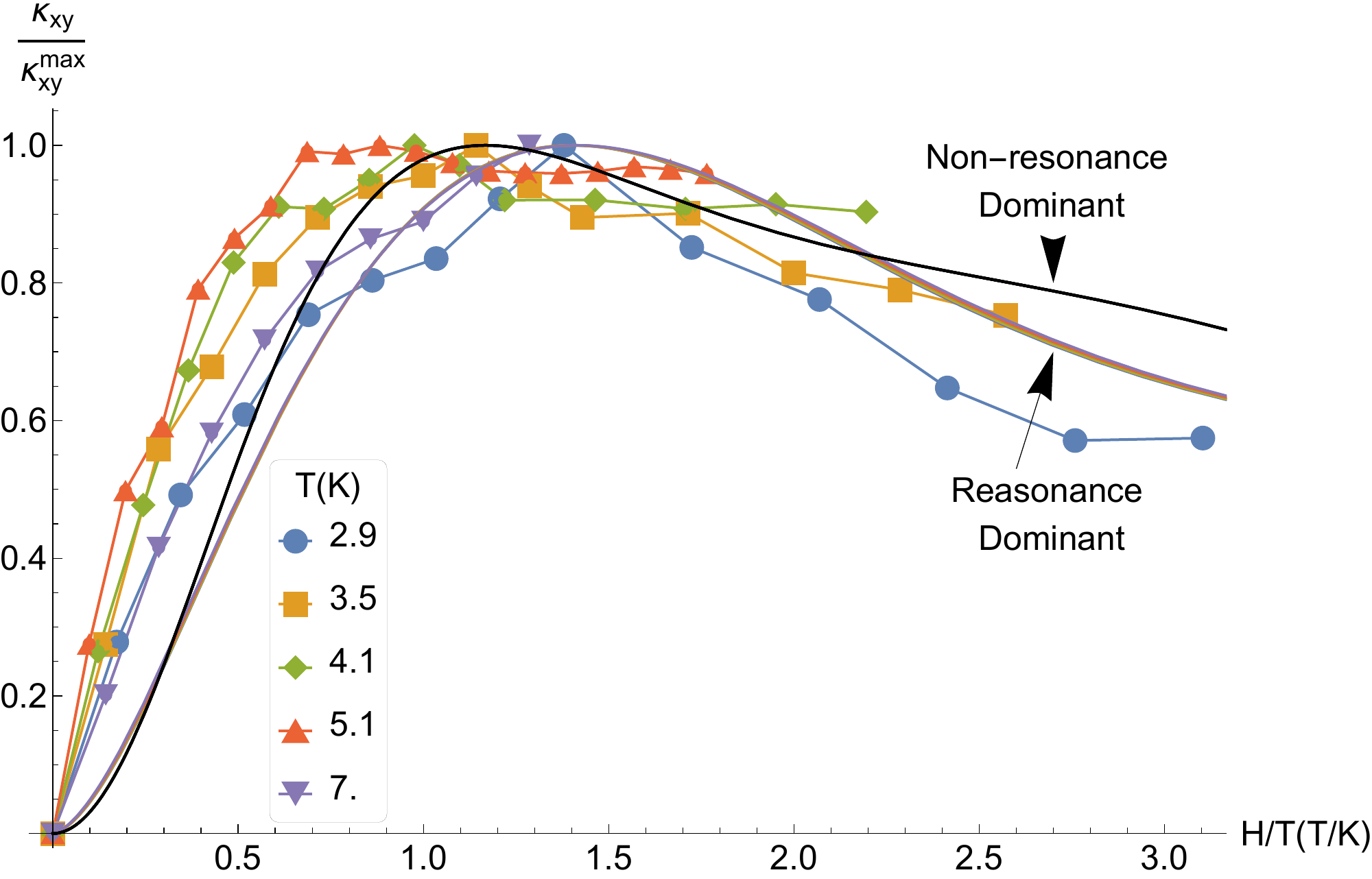}
  \caption{Similar to Fig.~\ref{fig:kxy001} but with field direction $H\parallel [111]$.}\label{fig:kxy111}
\end{figure}

We first consider the case with magnetic field $H\parallel[001]$ direction of the crystal. In this case, all the sublattices are equivalent and $\Delta=\frac{2}{\sqrt{3}}gJ\mu_B H$ \cite{Book_SpinIce,Uehara2022} (see also Appendix.~\ref{subsec:latticestructure}). In Figs.~\ref{fig:kxxT} and \ref{fig:MTC001}, we show a joint fitting of the zero-field $\kappa_{xx}$ data and the MTC data.
The Debye-Callaway model can reasonably approximate $\kappa_{xx}(H=0,T)$ as in Fig.~\ref{fig:kxxT}. We also see that after removing pseudospin-phonon coupling by setting $R=0$, the phonon peak reappears. Our simple model also captures the qualitative behavior of MTC (Fig.~\ref{fig:MTC001}), including a decrease at small $H$ and then  increase at larger $H$. However, the strength of resonant scattering $R$ seems to be overestimated (Appendix.~\ref{app:fitting}). From the fitting we extracted $gJ=3.42$, which is larger than the experimental value $gJ_\text{expr}=2.68$ \cite{Machida2005}, probably because our model use an oversimplified phonon band structure. We further use our extracted $gJ$ value to compute the thermal Hall effect and compare with the experimental data in Fig.~\ref{fig:kxy001}.
Following Ref.~\cite{Uehara2022},
we plot the normalized thermal Hall conductivity $\bar{\kappa}_{xy}(H,T)=\kappa_{xy}(T,H)/\max_H \left\{\kappa_{xy}(T,H)\right\}$. Notice that we have only performed fitting in $\kappa_{xx}$ and no fitting is done in $\kappa_{xy}$. Our model naturally explains the $H/T$ scaling collapse in $\bar{\kappa}_{xy}$ and reproduce the correct qualitative behavior. Quantitatively, the curve calculated from fitting parameters (colored curves) didn't capture the correct position of the peak. In account of the fact that the $\kappa_{xx}$ fitting overestimates resonant scattering, we also compared the data to the case where resonant scattering is subdominant, and found a better agreement (black curve).

Moving to the case of $H\parallel[111]$ direction, there are two kinds of pseudospins whose contributions add up to $\kappa_{xy}$. The first has $\Delta_1=2 gJ \mu_B H$ and occupies one sublattice, while the second has $\Delta_2=\frac{2}{3}gJ \mu_B H$ and occupies three sublattices (Appendix.~\ref{app:frame}). The prefactor $K_a$ of these two sublattices are equal (Appendix.~\ref{app:frame}).
  To obtain a stable fit, we fix their scattering strengths \eqref{eq:Gammaph} to be at the ratio $R_2/R_1=3/1$ by stoichiometry, and obtained $gJ=2.38$. We also found a larger deviation with the zero field $\kappa_{xx}$ data (not shown). Our computation for $\bar{\kappa}_{xy}$ is shown in Fig.~\ref{fig:kxy111}, which qualitatively reproduced a peak and a slow decay thereafter. Similar to the $[001]$ case, the non-resonant dominant curve agrees better with the data.

  A nice feature of the side-jump formalism is that the Hall angle is independent of the overall phonon mean-free path. In Appendix.~\ref{app:hallangle} we evaluate the Hall angles at $T\approx 20 {\rm K}$ and magnetic field $H=9{\rm T}$. The experimental measured Hall angles $0.4-0.8\times 10^{-3}$ near $20$K and $9$T is consistent with our previous estimated coupling constants.

\section{Discussion} Our simple paramagnetic model of phonon thermal Hall effect correctly captures some qualitative features of experimental data of Ref.~\cite{Uehara2022}. For the MTC, we have obtained an initial suppression due to the resonant scattering and later an enhancement. For $\bar{\kappa}_{xy}$, we obtained qualitatively correct shapes under magnetic field in both the $[001]$ and $[111]$ directions. In addition, our model naturally explains the approximate $H/T$ scaling of $\bar{\kappa}_{xy}$ in the experiment.  We also note that our model is oversimplified in several aspects, which we comment below:

(a) Currently, we have fitted for $\Gamma_\text{eff}$ assuming all three phonon bands are identical and isotropic. The situation might improve by including realistic phonon band structure and anisotropic scattering based on first principle calculation.

(b) In Fig.~\ref{fig:MTC001}, the strength of the resonant scattering in MTC is overestimated at intermediate field. One way to remedy the issue is to introduce quantum fluctuation \cite{Nakatsuji2006,Machida2010} or disorder. While incorporating the quantum effects and disorder is beyond the scope of the current work, we expect the energy spectrum of the pseudospin system is now broadened, and it causes the resonant singularity in $\Gamma_a(\omega)$ to be smeared. Therefore, the resonant scattering contribution to $\Gamma_\text{eff}$ is weakened, which reduces the suppression of $\kappa_{xx}$ at intermediate field.

(c) The same broadening of the pseudospin spectrum also has effect on $\bar{\kappa}_{xy}$ because fine details of $\bar{\kappa}_{xy}$ depends on the phonon scattering rate $\Gamma_\text{eff}$.  First, because the singularity in $\Gamma_\text{eff}$ has been smeared, we expect the non-resonant scattering to be more important in $\Gamma_\text{eff}$, and the resulting $\kappa_{xy}$ will look more like the black curves in Figs.~\ref{fig:kxy001} and \ref{fig:kxy111} which agree better with experiment.  Second, $\kappa_{xy}$ will have to be averaged over $\Delta$ according to the broadened energy spectrum. The effect of such an averaging process on $\kappa_{xy}$ depends on the detail of the quantum effects or disorder, and we defer it to future investigation.

(d) In Fig.~\ref{fig:MTC111}, the strength of the resonant scattering is underestimated. The situation here is more complicated than Fig.~\ref{fig:MTC001} because there are two types of pseudospins with different energy levels, and the shapes of the MTC curves depend sensitively on the ratio $R_2/R_1$ of the scattering strengths which, without further knowledge, we have set to $R_2/R_1=3$ based on stoichiometry. We expect the agreement to be improved if more microscopic information was available to determine $R_2/R_1$.

Phonon thermal Hall effect has also been observed in another non-Kramers pyrochlore compound $\rm{Tb}_2\rm{Ti}_2\rm{O}_7$ and its doped decendents \cite{Hirschberger2015,Hirokane2019}. We don't expect our theory to apply there because $\rm{Tb}^{3+}$ has a very soft CEF excited doublet at $19$K \cite{Gingras2000} so a two-level description breaks down. In this case we expect the skew-scattering mechanism to also be important.

{\it Note added:} After completion of the manuscript, we noticed a measurement of thermal Hall effect in $\rm{Pr}_2\rm{Zr_2}\rm{O}_7$ \cite{Chu2023} which is a sister compound of $\rm{Pr}_2\rm{Zr_2}\rm{O}_7$ and we expect our theory to also apply there.

\begin{acknowledgments}
We thank Jing-Yuan Chen, Jonathan Hallen, Yong Baek Kim, and Xiao-Qi Sun for helpful discussions. We thank Satoru Nakatsuji and Machida Yo for discussion and sharing experimental data. We thank Darshan G. Joshi and Subir Sachdev for discussion and collaboration on related projects. We also thank the support of KITP graduate fellowship, in part by the National Science Foundation under Grant No. NSF PHY-1748958. This work was partially performed at the Aspen Center for Physics, which is supported by National Science Foundation grant PHY-2210452. The participation of H.G. at the Aspen Center for Physics was supported by the Simons Foundation.
\end{acknowledgments}

\appendix

\newcommand{\rs}{{\rm s}}
\newcommand{\rK}{{\rm K}}
\begin{widetext}
\section{Derivation of Resonant Phonon Scattering Rate (Eq.\eqref{eq:Gammaph})}\label{app:resscat}

\subsection{Phonon diagonalization}

Following previous work \cite{Guo2022a}, the phonon Hamiltonian can be written as
\begin{equation}\label{eq:H=zhz}
  H_\text{ph}=\frac{1}{2}\sum_{ab}\zeta_a h_{ab} \zeta_b\,,
\end{equation} where $\zeta_a=\zeta_p^I=(u_p^i,\pi_p^i)$ encapsulates both ion displacement and momentum. We use 6-component capital indices $I,J,\dots$ to group 3-component cartesian indices $i,j,\dots$ of displacement $u^{i}_{p}$ and momentum $\pi^{i}_p$ together. $p,q,\dots$ denote lattice sites.
The $\zeta$'s satisfy the canonical commutation relation
\begin{equation}\label{eq:zz=iJ}
  [\zeta_a,\zeta_b]=iJ_{ab}\,.
\end{equation}

The band-diagonal basis is found by diagonalizing $iJh$:
\begin{equation}\label{}
  iJhM=M\ce\,,
\end{equation} where $\ce$ is a diagonal matrix.

For the acoustic phonon Hamiltonian given in the main text, the $h$ matrix and $J$ matrix have the following momentum space representation
\begin{equation}\label{eq:hk}
  h^{IJ}(k)=\begin{pmatrix}
         C^{ij}(k) & 0 \\
         0 & I_3
       \end{pmatrix}\,,
\end{equation}
\begin{equation}\label{eq:Jk}
  J^{IJ}(k)=J^{IJ}=\begin{pmatrix}
              0  & I_3 \\
              I_3 & 0
            \end{pmatrix}\,,
\end{equation} where $C^{ij}(k)$ is given by Eq.(2) in the main text, and $I_3$ denotes 3 by 3 identity matrix.

The Hamiltonian can now be diagonalized for each individual $k$. An explicit solution for the matrix $M(k)$ is
    \begin{equation}\label{}
    \begin{split}
      M(k)&=\begin{pmatrix}
             \begin{pmatrix}
               \frac{{e}^1_k}{\sqrt{m \omega^1_k}} & \frac{{e}^2_k}{\sqrt{m \omega^2_k}} & \frac{{e}^3_k}{\sqrt{m\omega^3_k}}
             \end{pmatrix} & 0 \\
             0 & \begin{pmatrix}
                   {e}^1_k\sqrt{m\omega^1_k} &  {e}^2_k\sqrt{m\omega^2_k} &  {e}^3_k\sqrt{m\omega^3_k}
                 \end{pmatrix}
           \end{pmatrix}\\
           &\times\left(\frac{1}{\sqrt{2}}\begin{pmatrix}
                    1 & 1 \\
                    -i & i
                  \end{pmatrix}\otimes I_3\right)\,.
    \end{split}
    \end{equation} The eigenvalues are
    \begin{equation}\label{}
      \ce=\mathrm{diag}(\omega^1_k,\omega^2_k,\omega^3_k,-\omega^1_k,-\omega^2_k,-\omega^3_k)\,.
    \end{equation}
    Here $e^a_k$ is a 3 by 1 column vector that describes the polarization vector of the $a$-th mode, and $\omega^a_k$ is the corresponding frequency. $I_3$ is the 3 by 3 identity matrix. We have one longitudinal mode $\omega^3_k=c_L k$ and two transverse modes $\omega^{1,2}_k=c_T k$.

    Following the same derivation in \cite{Guo2022a}, the phonon Green's function $D^{IJ}_{pq}(\tau)=-T_\tau\braket{\zeta_p^{I}(\tau)\zeta_q^J(0)}$ is found to be
\begin{equation}\label{eq:D}
  D(i\omega,k)=\frac{1}{i\omega-iJh(k)}iJ\,,
\end{equation} in Matsubara frequencies and matrix notation, in absence of interaction. Here $h$ and $J$ are matrices as defined in \eqref{eq:hk} amd \eqref{eq:Jk}. The phonon propagator can be diagonalized as
\begin{equation}\label{eq:tD}
  \tilde{D}^{ab}(i\omega,k)=\left[M^{-1}(k)D(i\omega,k)\left(iJ\right)^{-1}M(k)\right]_{ab}=\frac{1}{i\omega-\ce_a(k)}\delta^{ab}\,.
\end{equation}
\subsection{Phonon-pseudospin coupling}

We rewrite the phonon-pseudospin coupling in the main text as (duplicated $I,\alpha$ indices are implicitly summed)
\begin{equation}\label{}
  H_\text{ph-spin}=\sum_{p,q}\zeta_p^I B^{I\alpha}_{pq}\sigma_q^\alpha\,,
\end{equation} and the fourier transform of $B$ (in lattice sites $pq$) is
\begin{equation}\label{eq:Bk}
  B^{I\alpha}(k)=-ik_l K_{li\alpha}\,.
\end{equation} Here $i$ is the components of $I$ which denotes lattice displacement. We will also need the transpose of $B$ which satisfies $B^T(k)=B(-k)^T$.

\subsection{Pseudospin Correlations}

  We now compute the correlation function between pseudospins. Since pseudospins are spatially uncorrelated, we will restrict to a single lattice site and drop the site indices. We use the Abrikosov fermion method \cite{Joshi2020}, to fractionalize the pseudospin operators into spinons as
  \begin{equation}\label{}
    \sigma^\mu=f^\dagger_\alpha \tau^{\mu}_{\alpha\beta} f_\beta\,,\quad f_\alpha^\dagger f_\alpha=1\,.
  \end{equation} Here $\tau^\mu$ are Pauli matrices (renamed to avoid confusion). To enforce the half-filling constraint, we insert a chemical potential $\lambda$ for the fermions, so the Hamiltonian becomes
  \begin{equation}\label{}
    H_\text{spin}=\lambda-\frac{\Delta}{2}\sigma^3\,.
  \end{equation}

  The Abrikosov prescription says that any gauge-invariant pseudospin correlator should be calculated as
  \begin{equation}\label{}
    \braket{\dots}=\lim_{\lambda\to\infty}\frac{\braket{\dots}_W}{\braket{f_\alpha^\dagger f_\alpha}_W}\,.
  \end{equation} Here $\braket{}_W$ is calculated using Wick's theorem of spinon $f$'s.

  The Green's function of spinon is
  \begin{equation}\label{}
    \calG_{\alpha\beta}(i\omega)=-\int_0^{\beta}\rd \tau e^{-i\omega \tau} T_\tau\braket{f_\alpha(\tau)f^\dagger_\beta(0)}=\frac{1}{i\omega_n-\lambda+\frac{\Delta}{2}\tau^3}\,.
  \end{equation}

  The pseudospin two-point correlation is
  \begin{equation}\label{}
    S^{(2),\alpha\beta}(\tau)=-T_\tau\braket{\sigma^{\alpha}(\tau)\sigma^{\beta}(0)}\,.
  \end{equation} This is equivalent to a one-loop diagram of $f$-spinons
  \begin{equation}\label{}
    S^{(2),\alpha\beta}(i\omega)=\lim_{\lambda\to\infty}\frac{T \sum_{i\Omega} \Tr\left[\tau^\alpha\calG(i\omega+i\Omega)\tau^\beta\calG(i\Omega)\right]}{n_F(\lambda+\Delta/2)+n_F(\lambda-\Delta/2)}\,.
  \end{equation} The $\lambda\to\infty$ limit is evaluated after the Matsubara summation.
  The result is
\begin{equation}\label{eq:S2res}
  S^{(2),\alpha\beta}(i\omega)=\frac{2\tanh\frac{\beta\Delta}{2}}{(i\omega)^2-\Delta^2}(\Delta \delta^{\alpha\beta}-\epsilon^{\alpha\beta}\omega)\,,
\end{equation} where $\alpha,\beta=1,2$, $\delta^{\alpha\beta}$ is the Kronecker delta and $\epsilon^{\alpha\beta}$ is the Levi-Civita tensor with $\epsilon^{12}=1$.

Next we move on to the four-point correlation
\begin{equation}\label{}
       S^{(4),\alpha_1\alpha_2\alpha_3\alpha_4}(i\omega_1,i\omega_2,i\omega_3,i\omega_4)=\int \rd \tau_1\rd\tau_2\rd \tau_3\rd \tau_4 T_\tau\braket{\sigma^{\alpha_1}(\tau_1)\sigma^{\alpha_2}(\tau_2)\sigma^{\alpha_3}(\tau_3)\sigma^{\alpha_4}(\tau_4)}\exp(\sum_{j=1}^{4} i\omega_j\tau_j)\,.
     \end{equation} Here $\sum_{j=1}^{4}\omega_j=0$. By definition this is totally symmetric under argument exchange. The correlator can again be calculated as one-loop diagrams of Abrikosov fermions:
     \begin{equation}\label{}
       Q^{\alpha_1\alpha_2\alpha_3\alpha_4}(i\omega_1,i\omega_2,i\omega_3,i\omega_4)=-T\sum_{\Omega}\Tr\left[\sigma^{\alpha_1}\calG(i\Omega+i\omega_1)\sigma^{\alpha_2}\calG(i\Omega+i\omega_1+i\omega_2)\sigma^{\alpha_3}\calG(i\Omega-i\omega_4)\sigma^{\alpha_4}\calG(i\Omega)\right]\,,
     \end{equation}
     \begin{equation}\label{eq:S4Q}
       S^{(4),\alpha_1\alpha_2\alpha_3\alpha_4}(i\omega_1,i\omega_2,i\omega_3,i\omega_4)=\lim_{\lambda\to \infty}\frac{1}{n_F(\lambda+\epsilon_0/2)+n_F(\lambda-\epsilon_0/2)}\frac{1}{4}\sum_{P\in S_{4}} Q^{\alpha_{P1}\alpha_{P2}\alpha_{P3}\alpha_{P4}}(i\omega_{P1},i\omega_{P2},i\omega_{P3},i\omega_{P4})\,.
     \end{equation} Here the summation $P$ is over all permutations over $\{1,2,3,4\}$. When $P$ is the cyclic permutation $(1234)$, the diagram is over counted, and hence the factor $1/4$. Eq.\eqref{eq:S4Q} can be evaluated on Mathematica, but the general result is not very convenient to present, and we will discuss a special case we need later.

\subsection{Phonon Self energy}

    According to Schwinger-Dyson equation, the interacting phonon Green's function can be expanded in terms of self energy as
\begin{equation}\label{eq:D_SD}
\begin{tikzpicture}
  \draw[line width=3pt] (-0.5,0)--(0.5,0);
\end{tikzpicture}
=
\begin{tikzpicture}
  \draw[] (-0.5,0)--(0.5,0);
\end{tikzpicture}
+
\begin{tikzpicture}[baseline=0]
  \draw (-1.5,0)--(-0.5,0);
  \draw (0,0) circle (0.5);
  \node at (0,0) {$\Pi$};
  \draw (0.5,0)--(1.5,0);
\end{tikzpicture}
+
\begin{tikzpicture}[baseline=0]
  \draw (-1.5,0)--(-0.5,0);
  \draw (0,0) circle (0.5);
  \node at (0,0) {$\Pi$};
  \draw (0.5,0)--(1.5,0);
  \node at (2,0) {$\Pi$};
  \draw (2,0) circle (0.5);
  \draw (2.5,0)--(3.5,0);
\end{tikzpicture}
+\dots\,.
\end{equation} Here a thick line represents full phonon Green's function and a think line represents the free phonon Green's function.

However, in a system with (pseudo)spin, the perturbative expansion doesn't fit into the above structure. Expansion of phonon Green's function to fourth order in pseudospin-phonon coupling yields
\begin{equation}\label{eq:D_K}
\begin{split}
  \begin{tikzpicture}
  \draw[line width=3pt] (-0.5,0)--(0.5,0);
\end{tikzpicture}
=&
\begin{tikzpicture}
  \draw[] (-0.5,0)--(0.5,0);
\end{tikzpicture}
+
\sum_p
\begin{tikzpicture}[baseline=0]
  \draw (-1.5,0)--(-0.5,0);
  \draw[dotted] (-0.5,0)--(0.5,0);
  \node[below] at (0,0) {$p$};
  \draw (0.5,0)--(1.5,0);
  \filldraw[black] (-0.5,0) circle (0.05);
  \filldraw[black] (0.5,0) circle (0.05);
\end{tikzpicture}
+\sum_{p\neq q}
\begin{tikzpicture}[baseline=0]
  \draw (-1.5,0)--(-0.5,0);
  \draw[dotted] (-0.5,0)--(0.5,0);
  \node[below] at (0,0) {$p$};
  \draw (0.5,0)--(1.5,0);
  \filldraw[black] (-0.5,0) circle (0.05);
  \filldraw[black] (0.5,0) circle (0.05);
  \draw (2.5,0)--(3.5,0);
  \draw[dotted] (1.5,0)--(2.5,0);
  \node[below] at (2,0) {$q$};
  \filldraw[black] (1.5,0) circle (0.05);
  \filldraw[black] (2.5,0) circle (0.05);
\end{tikzpicture}\\
&+\sum_p
\begin{tikzpicture}[baseline=10]
  \draw (-1.5,0)--(-0.5,0);
  \node[below] at (0,0) {$p$};
  \draw (0.5,0)--(1.5,0);
  \filldraw[black] (-0.5,0) circle (0.05);
  \filldraw[black] (0.5,0) circle (0.05);
  \filldraw[black] (-0.5,1) circle (0.05);
  \filldraw[black] (0.5,1) circle (0.05);
  \filldraw[dotted,pattern=north east lines] (-0.5,0) rectangle (0.5,1);
  \draw (-0.5,1)..controls (0,1.5)..(0.5,1);
\end{tikzpicture}
+\dots\,.
\end{split}
\end{equation} Here a dotted line with letter $p$ represents spin two-point function at site $p$, and a black circle represents a coupling vertex $B^{I\alpha}$ and $K_{ij\alpha}$. Because the pseudospin doesn't satisfy Wick's theorem, there is an irreducible four-point correlation at each site $p$. Comparing Eqs.\eqref{eq:D_SD} and \eqref{eq:D_K}, we find to fourth order in $K_{ij\alpha}$:
\begin{equation}\label{}
\begin{tikzpicture}[baseline=0]
  \draw (0,0) circle (0.5);
  \node at (0,0) {$\Pi$};
\end{tikzpicture}=
\sum_p
\begin{tikzpicture}[baseline=0]
  \filldraw[black] (-0.5,0) circle (0.05);
  \filldraw[black] (0.5,0) circle (0.05);
  \draw[dotted] (-0.5,0)--(0.5,0);
  \node[below] at (0,0) {$p$};
\end{tikzpicture}
+\sum_p\left[
\begin{tikzpicture}[baseline=10]
\node[below] at (0,0) {$p$};
  \filldraw[black] (-0.5,0) circle (0.05);
  \filldraw[black] (0.5,0) circle (0.05);
  \filldraw[black] (-0.5,1) circle (0.05);
  \filldraw[black] (0.5,1) circle (0.05);
  \filldraw[dotted,pattern=north east lines] (-0.5,0) rectangle (0.5,1);
  \draw (-0.5,1)..controls (0,1.5)..(0.5,1);
\end{tikzpicture}-
\begin{tikzpicture}[baseline=0]
  \node[below] at (0,0) {$p$};
  \draw[dotted] (-0.5,0)--(0.5,0);
  \draw (0.5,0)--(1.5,0);
  \filldraw[black] (-0.5,0) circle (0.05);
  \filldraw[black] (0.5,0) circle (0.05);
  \draw[dotted] (1.5,0)--(2.5,0);
  \node[below] at (2,0) {$p$};
  \filldraw[black] (1.5,0) circle (0.05);
  \filldraw[black] (2.5,0) circle (0.05);
\end{tikzpicture}
\right]\,.
\end{equation} Translating the diagrams to equations, at second order we have
\begin{equation}\label{}
       \Pi^{(2)}(i\omega,k)=M^{-1}(k)(iJ)\tilde{\Pi}^{(2)}(i\omega,k)M(k)\,.
\end{equation} Here $\Pi^{(2)}$ means second order phonon self energy in band diagonal basis, and $\tilde{\Pi}^{(2)}$ is in the $\zeta$ basis, which can be written in matrix notation
\begin{equation}\label{}
  \tilde{\Pi}^{(2)}(i\omega,k)=B(k)S^{(2)}(i\omega)B^T(k)\,.
\end{equation} Here $B(k)$ is given by Eq.\eqref{eq:Bk}, $B^T(k)=\left[B(-k)\right]^T$ and $S^{(2)}$ is given by \eqref{eq:S2res}.

Continuing to real frequency by $i\omega\to \omega+i0$, we find that $\Im \Pi_R^{(2)}(\omega)\propto \delta(\omega^2-\Delta^2)$ only contains delta functions. This doesn't contribute to longitudinal transport because  $\Im \Pi_R$ appears in the denominator of phonon mean-free path, and a delta function in the denominator only suppress the integral at a zero-measure set of points.

Now moving to fourth order, we again have
\begin{equation}\label{eq:Pi4=tPi4}
  \Pi^{(4)}(i\omega,k)=M^{-1}(k)(iJ)\tilde{\Pi}^{(4)}(i\omega,k)M(k)\,,
\end{equation} where
\begin{equation}\label{eq:Pi4_1}
\begin{split}
       \tilde{\Pi}^{(4),IJ}(i\omega,k)=&\frac{1}{2}T\sum_{i\nu}\int \frac{\rd^3 q}{(2\pi)^3}B^{I \mu}(k)B^{J \nu}(-k) B^{K \rho}(-q)D^{KL}(i\nu,q)B^{L\lambda}(q)\\
       &\times\left[S^{(4),\mu\nu\rho \lambda}(i \omega,-i\omega,-i\nu,i\nu)-\beta S^{(2),\mu\rho}(i\omega)S^{(2),\lambda\nu}(i\omega)-\beta S^{(2),\mu\lambda}(i\omega)S^{(2),\rho\nu}(i\omega)\right]\,.
\end{split}
\end{equation} Here $1/2$ is a symmetry factor. According to optical theorem, the imaginary part of $\Pi^{(4)}$ must be factorizable into the norm squared of a second order on-shell process. By energy conservation the intermediate phonon must also have frequency $\nu=\omega$. This motivates us to look for delta-function singularities in $S^{(4)}$. Indeed, the Matsubara summation of Eq.\eqref{eq:S4Q} does yield different result when $\nu$ is set to $\pm \omega$ before the summation. We find
  \begin{equation}\label{eq:S4delta}
  S^{(4)}(i\omega,-i\omega,-i\nu,i\nu)=S^{(4)}_{reg}+S_+^{(4)}\beta \delta_{\omega,\nu}+S_-^{(4)}\beta \delta_{\omega,-\nu}.
\end{equation} Here $S^{(4)}_{reg}$ is a regular piece that doesn't contribute to imaginary part. The part that does contribute to imaginary part are the two delta function pieces, which are
\begin{equation}\label{eq:S4p}
\begin{split}
      S^{(4),\alpha_1 \alpha_2 \alpha_3\alpha_4}_{+}(i\omega)=&\frac{4}{(\Delta^2+\omega^2)^2}\left(\Delta  {\delta }_{\alpha _1\alpha _3}-\omega  {\epsilon }_{\alpha _1\alpha _3}\right)\\
       &\times\left(\Delta  {\delta }_{\alpha _2\alpha _4}+\omega  {\epsilon }_{\alpha _2\alpha _4}\right)\,,
\end{split}
\end{equation}
\begin{equation}\label{eq:S4m}
\begin{split}
   S^{(4),\alpha_1 \alpha_2 \alpha_3\alpha_4}_{-}(i\omega)=&\frac{4}{(\Delta^2+\omega^2)^2}\left(\Delta  {\delta }_{\alpha _1\alpha _4}-\omega  {\epsilon }_{\alpha _1\alpha _4}\right)\\
    &\times\left(\Delta  {\delta }_{\alpha _2\alpha _3}+\omega  {\epsilon }_{\alpha _2\alpha _3}\right)\,.\,
\end{split}
\end{equation} We note the above results have a factorized structure similar to $S^{(2)}S^{(2)}$ as anticipated from optical theorem.

Finally, we continue $i\omega\to \omega+i0$ in \eqref{eq:Pi4=tPi4} and compute the imaginary part of the diagonal components $-2\Im \Pi_R^{(aa)}(\omega)$. The imaginary part effectively converts the phonon Green's function in Eq.\eqref{eq:Pi4_1} into a delta function which pins the intermediate phonon to be on-shell. We assume phonon to be a good quasiparticle so we set external momentum to be on-shell $\omega_k=\omega$.  We also assume that other scattering mechanism of phonon effectively average the phonon momentum over different directions, so we can simplify the result. Finally, due to different normalizations of Fourier transforms on the lattice and in the continuum, we multiply the result by $a^3$ (unit cell volume) to restore the correct dimensionality, yielding Eq.(5) of the main text. Eq.(5) has three characteristic behaviors: (1) When $\omega\ll\Delta$, it is proportional to $\omega^4$, consistent with Rayleigh scattering; (2) When $\omega\approx \Delta$, there is a resonance; (3) When $\Delta\gg T$, the pseudospins are ordered and the scattering is exponentially suppressed, in accordance with Bloch's theorem.

\subsection{Validity of Perturbation theory}

In this part we argue the validity of the perturbation theory. First, the pseudospin physics is not dramatically altered by coupling to phonons. The problem of a single spin embedded in a bosonic bath has been studied before in Refs.~\cite{Sachdev1999b} and \cite{Vojta2000a}, and it was shown that direct coupling between a bosonic field and the spin is marginal in $3D$. For goldstone phonons, there is an additional derivative in the coupling and it renders the coupling irrelevant. Therefore perturbation theory is valid for the spin sector.

For the phonon sector, we need to compare the typical interaction energy $U$ with the energy of the resonant phonon, i.e. $\Delta$.  Using phonon mode expansion $u\sim(a+a^\dagger)/\sqrt{m\omega_k}$, at resonance we have $U\sim K \sqrt{\Delta/(mc^2)}$. Here $K$ is the typical coupling constant, $m$ is the typical ion mass and $c$ is typical speed of sound. The small parameter that justifies perturbation theory is $\alpha=U/\Delta\sim K/\sqrt{mc^2\Delta}$. Considering  $\rm{Pr}_2\rm{Ir}_2\rm{O}_7$, the typical coupling $K$ is estimated below in Sec.~\ref{sec:pointcharge}, and we have the average ion mass $m=23.8{\rm u}$ and typical sound velocity $c=3424{\rm m/s}$. Gathering these data, we found $\alpha\sim 1$ when the resonance energy $\Delta$ is of order $1{\rm K}$, and becomes smaller as $\Delta$ increases. This energy scale can be achieved by magnetic field less than $1{\rm T}$, so we expect $\alpha<1$ in a wide range of magnetic fields applied in the experiment \cite{Uehara2022}. Given that $\alpha<1$, we can follow arguments in \cite{Sun2022} to show that the perturbation theory converges due to fact that phonons are Goldstone modes.

\section{Derivation of spin-phonon coupling Eq.\eqref{eq:Hspin-ph}}\label{app:spin-ph-cpl}

    It has been shown in the literature \cite{Onoda2011,Rau2015} that the effective pseudospin-1/2 degrees of freedom in $\rm{Pr}_2\rm{Ir}_2\rm{O}_7$ arise from the splitting of the ${}^3H_4$ atomic state by crystal electric field (CEF) into (primarily) $M_J=\pm 4$ non-Kramers doublets. The pseudospin Hamiltonian can be derived by considering the superexchange interaction due to hybridization of $\rm{Pr}$ $4f$ electrons with the $p$ orbitals of nearby oxygen atoms. The above interactions depend on the interatomic distances and  therefore the system can be coupled to phonons. Since in $\rm{Pr}_2\rm{Ir}_2\rm{O}_7$ the CEF field interactions are strong (the first CEF excited state is at $168$K), we only consider the pseudospin-phonon coupling due to CEF effects.

  \subsection{Lattice Structure}
  \label{subsec:latticestructure}

    The $\rm{Pr}_2\rm{Ir}_2\rm{O}_7$ crystal can be embedded into a cubic lattice structure. The $\rm{Pr}$ atoms sit on vertices of corner sharing tetrahedra. There are two types of tetrahedra (up and down) where each tetrahedron connects to four of different types. In a cubic unit cell, there are 8 tetrahedra (4 up and 4 down) and 16 $\rm{Pr}$ atoms. The centers of the up and down tetrahedra form two interpenetrating fcc lattices.

    The specify the positions of the $\rm{Pr}$ atoms, we follow \cite{Onoda2011}, by starting with a fcc lattice of oxygen atoms
    \begin{equation}\label{}
      \vec{R}=\frac{a}{2}\left(n_2+n_3,n_1+n_3,n_1+n_2\right)\,.
    \end{equation} These sites form the center of one type of tetrahedra. The other set of fcc lattice sits at $\vec{R}'=\vec{R}-\frac{a}{4}(1,1,1)$. Since each $\rm{Pr}$ atom is shared by both types of tetrahedra, to enumerate the $\rm{Pr}$ atoms it is sufficient to use only the up tetrahedra, centered at $\vec{R}$.

    The $\rm{Pr}$ form 4 sublattices corresponding to the four vertices of a tetrahedron. We list the four sublattices sitting at $\vec{R}+\vec{a}_i$ as well as its local coordinate frame $(\vec{x}_i,\vec{y}_i,\vec{z}_i)$ ($i=0,1,2,3$):
    \begin{table}[hb!]
      \centering
      \begin{tabular}{|c|c|c|c|c|}
      \hline
       $i$ & 0 & 1 & 2 & 3 \\ \hline
      $\vec{a}_i$ & $-\frac{a}{8}(1,1,1)$ & $\frac{a}{8}(-1,1,1)$ & $\frac{a}{8}(1,-1,1)$ & $\frac{a}{8}(1,1,-1)$ \\ \hline
      $\vec{x}_i$ & $\frac{1}{\sqrt{6}}(1,1,-2)$ & $\frac{1}{\sqrt{6}}(1,-1,2)$ & $\frac{1}{\sqrt{6}}(-1,1,2)$ & $\frac{1}{\sqrt{6}}(-1,-1,-2)$ \\ \hline
      $\vec{y}_i$ & $\frac{1}{\sqrt{2}}(-1,1,0)$ & $\frac{1}{\sqrt{2}}(-1,-1,0)$ & $\frac{1}{\sqrt{2}}(1,1,0)$ & $\frac{1 }{\sqrt{2}}(1,-1,0)$ \\ \hline
      $\vec{z}_i$ & $\frac{1}{\sqrt{3}}(1,1,1)$ & $\frac{1}{\sqrt{3}}(1,-1,-1)$ & $\frac{1}{\sqrt{3}}(-1,1,-1)$ & $\frac{1}{\sqrt{3}}(-1,-1,1)$ \\
      \hline
    \end{tabular}
      \caption{Positions and local coordinate frames of four sublattices}\label{tab:frame}
    \end{table}

    Going into one of the local frames, the ionic environment of the $\rm{Pr}$ atom consists of ($\hat{x},\hat{y},\hat{z}$ below denote a particular set of $\vec{x}_i$, $\vec{y}_i$, $\vec{z}_i$ in Table.~\ref{tab:frame})
    \begin{enumerate}
      \item Two O1 atoms at $\pm\frac{\sqrt{3}}{8}a\hat{z}$.
      \item Six transition metal atoms TM at $C_6^n\frac{a}{2}\hat{y}$, $n=0,1,\dots,5$ where $C_6$ denote sixfold rotation along the $\hat{z}$ axis.
      \item Six O2 atoms at $\pm C_3^n \left(\sqrt{2}\left(\frac{1}{8}-\eta\right)\hat{x}+\eta \hat{z}\right)$, $n=0,1,2$ where $C_3$ is the threefold rotation along $\hat{z}$ and $\eta=0.0271$ for $\rm{Pr}_2\rm{Ir}_2\rm{O}_7$.
    \end{enumerate}

    The ionic environment is invariant under $D_{3d}$ group, which is generated by $C_{3}$ along $\hat{z}$ axis, $C_2$ along $\hat{y}$ axis and inversion $I$.

  \subsection{Symmetry analysis}

  In this part we derive the form of  pseudospin-phonon coupling in the absence of external magnetic field. The relevant symmetry group is $D_{3d}$ described above, and the fundamental representation of the generators are
  \begin{eqnarray}
    C_3 &=& \left(
\begin{array}{ccc}
 -\frac{1}{2} & -\frac{\sqrt{3}}{2} & 0 \\
 \frac{\sqrt{3}}{2} & -\frac{1}{2} & 0 \\
 0 & 0 & 1 \\
\end{array}
\right)\,, \label{eq:C3mat}\\
    C_2 &=& \left(
\begin{array}{ccc}
 -1 & 0 & 0 \\
 0 & 1 & 0 \\
 0 & 0 & -1 \\
\end{array}
\right)\,,\label{eq:C2mat} \\
    I &=& \left(
\begin{array}{ccc}
 -1 & 0 & 0 \\
 0 & -1 & 0 \\
 0 & 0 & -1 \\
\end{array}
\right)\,. \label{eq:Imat}
  \end{eqnarray}

  Since the ground state doublet is non-Kramers, $\sigma_x$ and $\sigma_y$ are even under time-reversal and allowed to couple to phonons. Microscopically, $\sigma_x$ and $\sigma_y$ correspond to quadrupole moments $J_x J_z$ and $J_yJ_z$ of the $\rm{Pr}$ atom respectively.

  We first derive the transformation of $\sigma_x$ and $\sigma_y$ under $D_{3d}$. Being pseudovectors, inversion $I$ acts as identity on them. We write $\sigma_x$ and $\sigma_y$ in terms of ground state doublet $\ket{\pm}_D$ as
  \begin{equation}\label{}
    \sigma_x=\sum_{\sigma=\pm} \ket{\sigma}_D \prescript{}{D}{\bra{\sigma}}\,, \quad \sigma_y=\sum_{\sigma=\pm}-i\sigma\ket{\sigma}_D \prescript{}{D}{\bra{\sigma}}\,.
  \end{equation}

  The exact expression of $\ket{\sigma}_D$ is given below in \eqref{eq:sigmaD} in terms of ${}^3H_4$ atomic states, but they transform the same way as $M_J=4$ states under $D_{3d}$:
  \begin{equation}\label{eq:Usigma}
    U(\Lambda)\ket{\sigma}_D=\sum_{\sigma'=\pm}\ket{\sigma'}_D \mathcal{D}^{j=4}_{4\sigma',4\sigma}(\Lambda)\,,
  \end{equation} where $\Lambda=C_3$ or $C_2$ and $\mathcal{D}$ is Wigner's $\mathcal{D}$-matrix of the corresponding rotations.

  Using \eqref{eq:Usigma} we can derive that
  \begin{equation}\label{eq:sigmatrans}
    U(\Lambda)\sigma_iU(\Lambda)^{\dagger}=\sum_{j}\sigma_j D^{\text{quad}}_{ji}(\Lambda)
  \end{equation}where
  \begin{eqnarray}
     D^\text{quad}\left(C_3\right)&=& \begin{pmatrix}
                                        -\frac{1}{2} & \frac{\sqrt{3}}{2} \\
                                        -\frac{\sqrt{3}}{2} & -\frac{1}{2}
                                      \end{pmatrix}\,, \\
    D^\text{quad}\left(C_2\right) &=& \begin{pmatrix}
                                        1 & 0 \\
                                        0 & -1
                                      \end{pmatrix}\,, \\
    D^\text{quad}\left(I\right) &=& \begin{pmatrix}
                                      1 & 0 \\
                                      0 & 1
                                    \end{pmatrix}\,.
  \end{eqnarray}

  The phonon strain operators $\partial_i u_j$ should transform as
  \begin{equation}\label{eq:dutrans}
    U(\Lambda)\partial_i u_j U(\Lambda)^\dagger=\sum_{i'j'}\left(\Lambda^{-1}\right)_{ii'} \partial_{i'}u_{j'} \Lambda_{j'j}\,,
  \end{equation} where $\Lambda$ denotes the generators defined in Eqs.\eqref{eq:C3mat}-\eqref{eq:Imat}. The transformations \eqref{eq:sigmatrans} and \eqref{eq:dutrans} look opposite to how a classical vector/tensor transforms, but it is required by the associativity of group action. Using \eqref{eq:sigmatrans} and \eqref{eq:dutrans}, we find the following invariants of $D_{3d}$ (same as Eq.\eqref{eq:Hspin-ph} with slightly different notation)
  \begin{equation}\label{eq:Hspin-ph2}
  \begin{split}
    H_\text{spin-ph}&=k_1\left[\sigma_x\left(\partial_x u_x-\partial_y u_y\right)+\sigma_y(\partial_xu_y+\partial_y u_x)\right]\\
               &+k_{2a}\left[\sigma_x \partial_x u_z-\sigma_y \partial_y u_z\right]\\
               &+k_{2b}\left[\sigma_x \partial_z u_x-\sigma_y \partial_z u_y\right]\,.
  \end{split}
  \end{equation} Here $k_1,k_{2a}$ and $k_{2b}$ are couplings invariant under $D_{3d}$. The coupling Hamiltonian \eqref{eq:Hspin-ph2} we have obtained is different from \cite{Patri2020}, due to using opposite transformations of the strain tensor.

  Under external magnetic field $\vec{h}$, the pseudospin $\sigma_z$ can linearly couple to $\vec{h}$. However, the thermal Hall effect is not sensitive to $\sigma_z$ couplings so we ignore this part.

  \subsection{Point charge analysis}\label{sec:pointcharge}

  In this part we review and extend the point charge analysis in \cite{Onoda2011} of the crystal electric field and derive microscopic expressions for the coupling constants in \eqref{eq:Hspin-ph} and \eqref{eq:Hspin-ph2}. We will work in the local frame of $\rm{Pr}$ atoms, so the $\rm{Pr}$ is at the origin.

  We consider the CEF Hamiltonian projected onto the ${}^3H_4$ ground state manifold of the $\rm{Pr}$ atom at origin, which reads
  \begin{equation}\label{eq:HCEF=VCEF}
    H_\text{CEF}=\sum_{m_l,m_l'=-3}^{3}V_\text{CEF}^{m_l,m_l'} \sum_{\sigma=\pm} f^\dagger_{m_l,\sigma} f_{m_l',\sigma}\,.
  \end{equation} Here $f_{m_l,\sigma}^\dagger$ is the creation operator of a $4f$ electron with magnetic quantum number $m_l$ and z-spin $\sigma/2$, and $V_\text{CEF}$ is the CEF potential experienced by a single electron.

  The ${}^3H_4$ atomic state contains two $4f$ electrons, the two electrons can be described by LS coupling scheme with  orbital angular momentum $L=5$,  spin angular momentum $S=1$, and total angular momentum $J=4$. We now express the eigenstates of $J_z$ using the creation operators and Clebsch-Gordon coefficients:
  \begin{equation}\label{}
  \begin{split}
    \ket{M_J}&=\sum_{M_L,M_S}C(LM_L,SM_S;JM_J)\ket{L,M_L;S,M_S}\\
            &=\sum_{M_L,M_S}C(LM_L,SM_S;JM_J)\sum_{m_l,m_l',\sigma,\sigma'}C(lm_l,l m_l';LM_L)C(\frac{1}{2}~\frac{\sigma}{2},\frac{1}{2}~\frac{\sigma'}{2};S M_S)\frac{1}{\sqrt{2}}f^\dagger_{m_l \sigma} f^\dagger_{m_l' \sigma'}\ket{0}\,,
  \end{split}
  \end{equation}where $C(j_1m_1,j_2m_2;jm)$ is the Clebsch-Gordon coefficient of merging transforming $\ket{j_1,m_1}\otimes \ket{j_2,m_2}$ to $\ket{j,m}$.

  The matrix element of the projected Hamiltonian is
  \begin{equation}\label{eq:HCEFMJMJp}
    H_\text{CEF}^{M_J,M_J'}=\braket{M_J|H_\text{CEF}|M_J'}\,.
  \end{equation} In \cite{Onoda2011}, it is found that the ground state of \eqref{eq:HCEFMJMJp} is given by the doublet
  \begin{equation}\label{eq:sigmaD}
    \ket{\sigma}_D=\alpha\ket{4\sigma}+\beta\sigma\ket{\sigma}-\gamma\ket{-2\sigma}\,,\quad \sigma=\pm1\,,
  \end{equation} where $\beta,\gamma$ are real numbers and $\alpha=\sqrt{1-\beta^2-\gamma^2}$. The best fit for neutron scattering data corresponds to $\beta=0.075,\gamma=3\beta$.

  In the presence of lattice distortions $\vec{u}(\vec{R})$, $H_\text{CEF}$ will depend on the strain $\partial_i u_j$. Applying the first order degenerate perturbation theory, the phonon-pseudospin coupling Hamiltonian is then given by
  \begin{equation}\label{eq:Hspinphss'}
    H_\text{spin-ph}=\sum_{\sigma,\sigma'}\ket{\sigma}_D\prescript{}{D}{\braket{\sigma|H^{(1)}_\text{CEF}(\partial u)|\sigma'}}_D \prescript{}{D}{\bra{\sigma'}}\,.
  \end{equation} Here $H^{(1)}_{CEF}(\partial u)$ denotes the part of the CEF Hamiltonian that is first order in the strains $\partial_i u_j$, which we compute below.

  In the presence of small lattice distortions $\vec{u}(\vec{R})$, the single particle CEF potential $V_\text{CEF}$ can be written in real space basis as
  \begin{equation}\label{}
    V_\text{CEF}(\vec{r})=-e\sum_{a}\frac{q_i}{\left|\vec{R}_a+\vec{u}(\vec{R}_a)-\vec{r}-\vec{u}(0)\right|}\,.
   \end{equation} Here the eletron charge is $-e$ and the ions are situated at positions $\left\{\vec{R}_a\right\}$ with charges $\left\{q_a\right\}$.
   The summation over $a$ runs through the three types of ions discussed in Sec.~\ref{subsec:latticestructure}, with effective charges $q_\text{O1}$, $q_\text{O2}$ and $q_\text{TM}$. Expanding to first order in the gradient $\nabla \vec{u}$, we obtain
   \begin{equation}\label{eq:V(1)}
     V^{(1)}_\text{CEF}(\vec{r})=\sum_a \frac{eq_a}{|\vec{R}_a-\vec{r}|^3}
     \partial_i u_j R_{ai}\left(R_{aj}-r_j\right)\,,
   \end{equation} where repeated Cartesian indices are summed.

   To obtain the matrix elements  $V^{(1),m,m'}_\text{CEF}$, we need to expand \eqref{eq:V(1)} into spherical Harmonics. An useful formula is
   \begin{equation}\label{eq:1/(R-r)^3}
     \frac{1}{|\vec{R}-\vec{r}|^3}=\frac{1}{R^3}\sum_{l=0}\frac{(r/R)^{l}}{1-(r/R)^2}4\pi\sum_{m=-l}^{l}(-1)^m Y_{l}^{-m}(\hat{r})Y_l^{m}(\hat{R})\,, \quad r<R\,,
   \end{equation} where $\vec{R}=R\hat{R}$, $\vec{r}=r\hat{r}$, $\hat{R}$ and $\hat{r}$ are unit vectors. Using \eqref{eq:1/(R-r)^3} and Clebsch-Gordon coefficients, we can expand the tensor $$\frac{R_{ai}(R_{aj}-r_j)}{|\vec{R}_a-\vec{r}|^3}$$ into a bilinear of the spherical Harmonics $Y_{l}^{m}(\hat{R}_a)$ and $Y_{l}^{m}(\hat{r})$. The matrix element $V^{(1),m,m'}_\text{CEF}$ is then given as
   \begin{equation}\label{eq:V1mm'}
     V^{(1),m_l,m_l'}_\text{CEF}=\sum_a eq_a\partial_i u_j\int \rd^2\hat{r} Y_{3}^{m}(\hat{r})^{*}\left\langle\frac{R_{ai}(R_{aj}-r_j)}{|\vec{R}_a-\vec{r}|^3}\right\rangle Y_{3}^{m'}(\hat{r})\,,
   \end{equation} where $\int \rd^2\hat{r}$ denotes the angular average of $\vec{r}$ and $\braket{\cdot}$ denotes the radial average.

   Finally, plugging the result of \eqref{eq:V1mm'} into \eqref{eq:HCEF=VCEF}, evaluating the 2 by 2 matrix  in \eqref{eq:Hspinphss'} and rewriting the result with Pauli matrices, we successfully recover the Hamiltonian \eqref{eq:Hspin-ph} and \eqref{eq:Hspin-ph2}, and the couplings are given by

    \begin{equation}\label{}
    \begin{split}
    k_1/e=&\frac{q_{\text{O2}}} {R_{\text{O2}}} \Bigg\{\frac{\braket{r^4}}{R_{\text{O2}}^4} \bigg[-\frac{675 \sqrt{7} \alpha  \gamma  \sin ^2\left(\theta _2\right)}{1936}-\frac{315}{484} \sqrt{7} \alpha  \gamma  \sin ^2\left(\theta _2\right) \cos \left(2 \theta _2\right)-\frac{945 \sqrt{7} \alpha  \gamma  \sin ^2\left(\theta _2\right) \cos \left(4 \theta _2\right)}{1936}\\
    &+\frac{675 \beta ^2 \sin ^2\left(\theta _2\right)}{1936}+\frac{315}{484} \beta ^2 \sin ^2\left(\theta _2\right) \cos \left(2 \theta _2\right)+\frac{945 \beta ^2 \sin ^2\left(\theta _2\right) \cos \left(4 \theta _2\right)}{1936}-\frac{315 \beta  \gamma  \sin \left(2 \theta _2\right) \sin ^2\left(\theta _2\right)}{242 \sqrt{2}}\\
    &-\frac{945 \beta  \gamma  \sin \left(4 \theta _2\right) \sin ^2\left(\theta _2\right)}{484 \sqrt{2}}-\frac{945}{968} \gamma ^2 \sin ^6\left(\theta _2\right)\bigg]\\
    &+\frac{\braket{r^6}}{R_{\text{O2}}^6} \bigg[\frac{357}{572} \sqrt{\frac{7}{2}} \alpha  \beta  \sin \left(2 \theta _2\right) \sin ^4\left(\theta _2\right)+\frac{51}{44} \sqrt{\frac{7}{2}} \alpha  \beta  \sin \left(2 \theta _2\right) \sin ^4\left(\theta _2\right) \cos \left(2 \theta _2\right)+\frac{14875 \sqrt{7} \alpha  \gamma  \sin ^2\left(\theta _2\right)}{50336}\\
    &+\frac{57375 \sqrt{7} \alpha  \gamma  \sin ^2\left(\theta _2\right) \cos \left(2 \theta _2\right)}{100672}+\frac{2295 \sqrt{7} \alpha  \gamma  \sin ^2\left(\theta _2\right) \cos \left(4 \theta _2\right)}{4576}+\frac{255}{704} \sqrt{7} \alpha  \gamma  \sin ^2\left(\theta _2\right) \cos \left(6 \theta _2\right)\\
    &+\frac{104125 \beta ^2 \sin ^2\left(\theta _2\right)}{201344}+\frac{401625 \beta ^2 \sin ^2\left(\theta _2\right) \cos \left(2 \theta _2\right)}{402688}+\frac{16065 \beta ^2 \sin ^2\left(\theta _2\right) \cos \left(4 \theta _2\right)}{18304}\\
    &+\frac{1785 \beta ^2 \sin ^2\left(\theta _2\right) \cos \left(6 \theta _2\right)}{2816}-\frac{116739 \beta  \gamma  \sin \left(2 \theta _2\right) \sin ^2\left(\theta _2\right)}{50336 \sqrt{2}}-\frac{3213 \beta  \gamma  \sin \left(2 \theta _2\right) \sin ^2\left(\theta _2\right) \cos \left(2 \theta _2\right)}{1144 \sqrt{2}}\\
    &-\frac{1071 \beta  \gamma  \sin \left(2 \theta _2\right) \sin ^2\left(\theta _2\right) \cos \left(4 \theta _2\right)}{352 \sqrt{2}}-\frac{357}{104} \gamma ^2 \sin ^6\left(\theta _2\right)-\frac{357}{88} \gamma ^2 \sin ^6\left(\theta _2\right) \cos \left(2 \theta _2\right)\bigg]\\
    &+\frac{\braket{r^2}}{R_{\text{O2}}^2} \bigg[-\frac{39}{275} \sqrt{7} \alpha  \gamma  \sin ^2\left(\theta _2\right)-\frac{13}{55} \sqrt{7} \alpha  \gamma  \sin ^2\left(\theta _2\right) \cos \left(2 \theta _2\right)-\frac{39}{110} \beta ^2 \sin ^2\left(\theta _2\right)-\frac{13}{22} \beta ^2 \sin ^2\left(\theta _2\right) \cos \left(2 \theta _2\right)\\
    &+\frac{117 \beta  \gamma  \sin \left(2 \theta _2\right) \sin ^2\left(\theta _2\right)}{55 \sqrt{2}}\bigg]\Bigg\}\\
    &+\frac{q_{\text{TM}}}{ R_{\text{TM}}} \Bigg\{\left(-\frac{425 \sqrt{7} \alpha  \gamma }{3146}-\frac{2975 \beta ^2}{12584}-\frac{357 \gamma ^2}{572}\right) \frac{\braket{r^6}}{R_{\text{TM}}^6}\\
    &+\left(-\frac{45}{242} \sqrt{7} \alpha  \gamma +\frac{45 \beta ^2}{242}+\frac{945 \gamma ^2}{968}\right) \frac{\braket{r^4}}{R_{\text{TM}}^4}+\left(\frac{26}{275} \sqrt{7} \alpha  \gamma +\frac{13 \beta ^2}{55}\right) \frac{\braket{r^2}}{R_{\text{TM}}^2}\Bigg\}\,,
    \end{split}
    \end{equation}
    \begin{equation}\label{}
    \begin{split}
       k_{2a}/e= & \frac{q_{\text{O2}}}{ R_{\text{O2}}} \Bigg\{\frac{\braket{r^6}}{R_{\text{O2}}^6} \Bigg[
       -\frac{357 \sqrt{\frac{7}{2}} \alpha  \beta  \sin ^2\left(\theta _2\right)}{1144}
       -\frac{153 \sqrt{\frac{7}{2}} \alpha  \beta  \sin ^2\left(\theta _2\right) \cos \left(2 \theta _2\right)}{2288}
       +\frac{765 \sqrt{\frac{7}{2}} \alpha  \beta  \sin ^2\left(\theta _2\right) \cos \left(4 \theta _2\right)}{1144}\\
       &-\frac{51}{176} \sqrt{\frac{7}{2}} \alpha  \beta  \sin ^2\left(\theta _2\right) \cos \left(6 \theta _2\right)-\frac{19125 \sqrt{7} \alpha  \gamma  \sin \left(2 \theta _2\right) \sin ^2\left(\theta _2\right)}{50336}-\frac{765 \sqrt{7} \alpha  \gamma  \sin \left(4 \theta _2\right) \sin ^2\left(\theta _2\right)}{1144}\\
       &-\frac{255}{352} \sqrt{7} \alpha  \gamma  \sin \left(6 \theta _2\right) \sin ^2\left(\theta _2\right)-\frac{133875 \beta ^2 \sin \left(2 \theta _2\right) \sin ^2\left(\theta _2\right)}{201344}-\frac{5355 \beta ^2 \sin \left(4 \theta _2\right) \sin ^2\left(\theta _2\right)}{4576}\\
       &-\frac{1785 \beta ^2 \sin \left(6 \theta _2\right) \sin ^2\left(\theta _2\right)}{1408}-\frac{62475 \beta  \gamma  \sin ^2\left(\theta _2\right)}{25168 \sqrt{2}}-\frac{240975 \beta  \gamma  \sin ^2\left(\theta _2\right) \cos \left(2 \theta _2\right)}{50336 \sqrt{2}}-\frac{9639 \beta  \gamma  \sin ^2\left(\theta _2\right) \cos \left(4 \theta _2\right)}{2288 \sqrt{2}}\\
       &-\frac{1071 \beta  \gamma  \sin ^2\left(\theta _2\right) \cos \left(6 \theta _2\right)}{352 \sqrt{2}}-\frac{5355 \gamma ^2 \sin \left(2 \theta _2\right) \sin ^2\left(\theta _2\right)}{9152}-\frac{1071 \gamma ^2 \sin \left(4 \theta _2\right) \sin ^2\left(\theta _2\right)}{2288}+\frac{357}{704} \gamma ^2 \sin \left(6 \theta _2\right) \sin ^2\left(\theta _2\right)\Bigg]\\
       &+\frac{\braket{r^4}}{R_{\text{O2}}^4} \Bigg[\frac{315}{484} \sqrt{7} \alpha  \gamma  \sin \left(2 \theta _2\right) \sin ^2\left(\theta _2\right)+\frac{945}{968} \sqrt{7} \alpha  \gamma  \sin \left(4 \theta _2\right) \sin ^2\left(\theta _2\right)-\frac{315}{484} \beta ^2 \sin \left(2 \theta _2\right) \sin ^2\left(\theta _2\right)\\
       &-\frac{945}{968} \beta ^2 \sin \left(4 \theta _2\right) \sin ^2\left(\theta _2\right)-\frac{675 \beta  \gamma  \sin ^2\left(\theta _2\right)}{242 \sqrt{2}}-\frac{315}{121} \sqrt{2} \beta  \gamma  \sin ^2\left(\theta _2\right) \cos \left(2 \theta _2\right)-\frac{945 \beta  \gamma  \sin ^2\left(\theta _2\right) \cos \left(4 \theta _2\right)}{242 \sqrt{2}}\\
       &-\frac{945 \gamma ^2 \sin \left(2 \theta _2\right) \sin ^2\left(\theta _2\right)}{1936}+\frac{945 \gamma ^2 \sin \left(4 \theta _2\right) \sin ^2\left(\theta _2\right)}{3872}\Bigg]\\
       &+\frac{\braket{r^2}}{R_{\text{O2}}^2} \left[\frac{26}{55} \sqrt{7} \alpha  \gamma  \sin \left(2 \theta _2\right) \sin ^2\left(\theta _2\right)+\frac{13}{11} \beta ^2 \sin \left(2 \theta _2\right) \sin ^2\left(\theta _2\right)+\frac{351}{275} \sqrt{2} \beta  \gamma  \sin ^2\left(\theta _2\right)+\frac{117}{55} \sqrt{2} \beta  \gamma  \sin ^2\left(\theta _2\right) \cos \left(2 \theta _2\right)\right]\Bigg\}\\
       &+\frac{q_{\text{TM}}}{ R_{\text{TM}}} \left(\left(\frac{1785 \beta  \gamma }{1573 \sqrt{2}}-\frac{51}{143} \sqrt{14} \alpha  \beta \right) \frac{\braket{r^6}}{R_{\text{TM}}^6}-\frac{90}{121} \sqrt{2} \beta  \gamma  \frac{\braket{r^4}}{R_{\text{TM}}^4}-\frac{234}{275} \sqrt{2} \beta  \gamma  \frac{\braket{r^2}}{R_{\text{TM}}^2}\right)
    \end{split}
    \end{equation}
    \begin{equation}\label{}
      \begin{split}
         k_{2b}/e & =\frac{q_{\text{O1}}}{ R_{\text{O1}}} \left(\frac{952 \sqrt{2} \beta  \gamma  \frac{\braket{r^6}}{R_{\text{O1}}^6}}{1573}+\frac{80}{121} \sqrt{2} \beta  \gamma  \frac{\braket{r^4}}{R_{\text{O1}}^4}-\frac{156}{275} \sqrt{2} \beta  \gamma  \frac{\braket{r^2}}{R_{\text{O1}}^2}\right)\\
         &+\frac{q_{\text{O2}}}{ R_{\text{O2}}} \Bigg\{\frac{\braket{r^4}}{R_{\text{O2}}^4} \Bigg[\frac{2415}{484} \sqrt{7} \alpha  \gamma  \sin ^3\left(\theta _2\right) \cos \left(\theta _2\right)+\frac{945}{484} \sqrt{7} \alpha  \gamma  \sin ^3\left(\theta _2\right) \cos \left(3 \theta _2\right)-\frac{2415}{484} \beta ^2 \sin ^3\left(\theta _2\right) \cos \left(\theta _2\right)\\
         &-\frac{945}{484} \beta ^2 \sin ^3\left(\theta _2\right) \cos \left(3 \theta _2\right)+\frac{435 \beta  \gamma  \cos ^2\left(\theta _2\right)}{242 \sqrt{2}}-\frac{105}{121} \sqrt{2} \beta  \gamma  \cos ^2\left(\theta _2\right) \cos \left(2 \theta _2\right)+\frac{945 \beta  \gamma  \cos ^2\left(\theta _2\right) \cos \left(4 \theta _2\right)}{242 \sqrt{2}}\\
         &+\frac{2415 \gamma ^2 \sin ^3\left(\theta _2\right) \cos \left(\theta _2\right)}{1936}+\frac{945 \gamma ^2 \sin ^3\left(\theta _2\right) \cos \left(3 \theta _2\right)}{1936}\Bigg]\\
         &+\frac{\braket{r^6}}{R_{\text{O2}}^6} \Bigg[-\frac{51}{11} \sqrt{14} \alpha  \beta  \sin ^6\left(\theta _2\right) \cos ^2\left(\theta _2\right)-\frac{87465 \sqrt{7} \alpha  \gamma  \sin ^3\left(\theta _2\right) \cos \left(\theta _2\right)}{12584}-\frac{10455 \sqrt{7} \alpha  \gamma  \sin ^3\left(\theta _2\right) \cos \left(3 \theta _2\right)}{2288}\\
         &-\frac{255}{176} \sqrt{7} \alpha  \gamma  \sin ^3\left(\theta _2\right) \cos \left(5 \theta _2\right)-\frac{612255 \beta ^2 \sin ^3\left(\theta _2\right) \cos \left(\theta _2\right)}{50336}-\frac{73185 \beta ^2 \sin ^3\left(\theta _2\right) \cos \left(3 \theta _2\right)}{9152}\\
         &-\frac{1785}{704} \beta ^2 \sin ^3\left(\theta _2\right) \cos \left(5 \theta _2\right)-\frac{18921 \beta  \gamma  \cos ^2\left(\theta _2\right)}{25168 \sqrt{2}}+\frac{138159 \beta  \gamma  \cos ^2\left(\theta _2\right) \cos \left(2 \theta _2\right)}{50336 \sqrt{2}}\\
         &-\frac{3213 \beta  \gamma  \cos ^2\left(\theta _2\right) \cos \left(4 \theta _2\right)}{2288 \sqrt{2}}+\frac{1071 \beta  \gamma  \cos ^2\left(\theta _2\right) \cos \left(6 \theta _2\right)}{352 \sqrt{2}}+\frac{122451 \gamma ^2 \sin ^3\left(\theta _2\right) \cos \left(\theta _2\right)}{25168}+\frac{14637 \gamma ^2 \sin ^3\left(\theta _2\right) \cos \left(3 \theta _2\right)}{4576}\\
         &+\frac{357}{352} \gamma ^2 \sin ^3\left(\theta _2\right) \cos \left(5 \theta _2\right)\Bigg]\\
         &+\frac{\braket{r^2}}{R_{\text{O2}}^2} \left(\frac{52}{55} \sqrt{7} \alpha  \gamma  \sin ^3\left(\theta _2\right) \cos \left(\theta _2\right)+\frac{26}{11} \beta ^2 \sin ^3\left(\theta _2\right) \cos \left(\theta _2\right)-\frac{234}{55} \sqrt{2} \beta  \gamma  \cos ^4\left(\theta _2\right)+\frac{702}{275} \sqrt{2} \beta  \gamma  \cos ^2\left(\theta _2\right)\right)\Bigg\}
      \end{split}
    \end{equation}
     Here $\theta_2=\arctan\left[\frac{\sqrt{2}}{\eta}\left(1/8-\eta\right)\right]$ is the polar angle of O2 oxygens in the $z>0$ half plane. To estimate of order of magnitude, we assume $\frac{eq}{R}$ is about $1$eV, as $R$ is about one order of magnitude larger than Bohr's radius. We also see that the couplings all involve at least one factor of $\beta$ or $\gamma$, which brings the estimate down by another order of magnitude, yielding a crude estimate from $10^2$K to $10^3$K  for the coupling constants. This is also the typical span the CEF spectrum \cite{Machida2005}.

\section{Thermal Hall effect (Derivation of Eq.\eqref{eq:kxyint})}\label{app:the}

In this section we review the computations of side-jump thermal Hall effect in Ref.~\cite{Guo2022a} and discuss how to include the feedback due to resonance. Additionally we also discuss the reference frame issues when the result is applied to pyrochlore crystal.

\subsection{Side-jump thermal Hall effect with feedback}

The side-jump thermal Hall effect can be computed using the Kubo formula as the magnetization correction is unimportant \cite{Guo2022a}. Following the procedures in Ref.~\cite{Guo2022a}, the side-jump thermal Hall effect can be written as
\begin{equation}\label{}
  \kappa_H=\kappa_H^{(0)}+\kappa_H^{(1a)}+\kappa_H^{(1b)}\,,
\end{equation}where
\begin{equation}\label{eq:kappaH0}
\begin{split}
  \kappa_H^{(0)}&=\frac{-\beta}{2\pi}\int \rd z (-n_B'(z))\Tr[V_f^{(0)}\tilde{D}_+\Pi_+\tilde{D}_+V_g^{(0)}\tilde{D}_-
  +V_f^{(0)}\tilde{D}_+V_g^{(0)}\tilde{D}_-{\Pi}_-\tilde{D}_-]\\
  &-(f\leftrightarrow g)\,,
\end{split}
\end{equation}
\begin{equation}\label{eq:kappaH1a_0}
\begin{split}
  \kappa_H^{(1a)}=\frac{\beta}{2\pi}&\int \rd z(-n_B'(z))\Tr\Bigg[\frac{[hJh,f]}{4}D_+\frac{[h,g]JB}{2}G_-+\frac{[h,f]JB}{2}G_+\frac{[hJh,g]}{4}D_-\Bigg]-(f\leftrightarrow g)\,,
\end{split}
\end{equation}
\begin{equation}\label{eq:kappaH1b_0}
\begin{split}
  \kappa_H^{(1b)}=\frac{\beta}{2\pi}&\int \rd z(-n_B'(z))\Tr\Bigg[\frac{[hJh,f]}{4}D_+h[JB,g]G_-+h[JB,f]G_+\frac{[hJh,g]}{4}D_-\Bigg]-(f\leftrightarrow g)\,.
\end{split}
\end{equation} Here $D$ and $\tilde{D}$ are the phonon Green's function in different basis as defined in Eqs.\eqref{eq:D} and \eqref{eq:tD}; $G=S^{(2)}B^T D$ where $S^{(2)}$ is the spin Green's function in Eq.\eqref{eq:S2res} and $B$ is defined by Eq.\eqref{eq:Bk}. The $\pm$ in the subscripts denote retarded/advanced Green's function.
 In Eqs.\eqref{eq:kappaH0}-\eqref{eq:kappaH1b_0}, the trace denotes sum over lattice site and cartesian component indices (or equivalently, sum over momentum and band indices), and the commutators are understood in a similar manner.
In Eq.\eqref{eq:kappaH0}
The vertex function is
\begin{equation}\label{eq:Vf0}
  V_f^{(0)}=\frac{M^{-1}[(iJh)^2,f]M}{4}=\frac{1}{4}\left([\ce^2,f]+[\ce^2,A_f]\right)\,,
\end{equation} where we have separated it into band-diagonal and band-off diagonal parts. The off-diagonal part is given by the connection $A_f$:
\begin{equation}\label{eq:Af}
  A_f=-M^{-1}[M,f]\,.
\end{equation} The commutator with $f,g$ are momentum derivatives, i.e.  $[H(k),f]=i\partial_{k_x}H(k)$, $[H(k),g]=i\partial_{k_y}H(k)$.

The side-jump contribution corresponds to the situation that one energy current vertex is intra-band and the other is inter-band \cite{Sinitsyn2007}. Since in Eqs.\eqref{eq:kappaH0}-\eqref{eq:kappaH1b_0}, there is at most one retarded or advanced propagator, such a contribution is easy to pick out. To proceed with the $z$-integral, we use the identity
\begin{equation}\label{}
  \tilde{D}_+^{a}(z)\tilde{D}_{-}^{a}(z)=\frac{i}{\Gamma_{a}(z)}\left[\tilde{D}_+^{a}(z)-\tilde{D}_-^{a}(z)\right]\,,
\end{equation}  and then we evaluate the integral using residue method by dropping the poles in $n_B'(z)$, which is equivalent to assuming phonons are well-defined quasiparticles $\Gamma_{a}(z)\ll T$ \cite{Abrikosov1963}. This procedure is tantamount to setting $z=\ce_a\pm i\Gamma_a/2$. The manipulations in the previous work \cite{Guo2022a} did not include this imaginary part of $z$, which is sufficient when the resonance is not dominant but fails in the regime when the resonance effect is strong.

 From this point, we follow the same algebra as \cite{Guo2022a} and arrive at the following semiclassical expression for thermal Hall effect
\begin{equation}\label{eq:kappaH_semiclassical}
\begin{split}
  \kappa_{xy}^{sj}=&\frac{1}{2}\sum_{a}\int \frac{\rd^3 k}{(2\pi)^3} \frac{(-\beta n_B'(\ce_a))}{\Gamma_a(k)}j^E_{\text{on-shell},x}j^E_{\text{side-jump},y}\\
  &-(x\leftrightarrow y)\,.
\end{split}
\end{equation} Here $a$ sums over phonon bands (positive and negative frequency modes count as two separate bands); $\ce_a=\ce_a(k)$ is the phonon dispersion of the $a$-th band; $\Gamma_a(k)$ is the phonon decay-rate and $n_B$ is the Bose distribution function. Eq.\eqref{eq:kappaH_semiclassical} resembles the solution of Boltzmann equation, with two kinds of energy currents. The first one is
\begin{equation}\label{}
  j^E_{\text{on-shell},x}=\ce_a c_{a,x}
\end{equation} which is the usual term of energy multiplied by velocity. The second energy current is
\begin{equation}\label{eq:jEsj}
  j^E_{\text{side-jump},y}= (c_{{\rm sj},y}^{aa}+\sum_{b\neq a}c_{{\rm sj},y}^{ba})\ce_a\,,
\end{equation} which takes the form of energy multiplied by side-jump velocities. The first term $c^{aa}_{{\rm sj}}$ is due to renormalization of phonon velocity due to the pseudospin-phonon coupling, given by
  \begin{equation}\label{eq:vsjaa}
  v_{\rm sj}^{aa}(g)=\left(M^{-1}[iJB,g](-i)\left[S_-\left(\ce_a-\frac{i}{2}\Gamma_a(k)\right)-S_+\left(\ce_a+\frac{i}{2}\Gamma_a(k)\right)\right]B^T M\right)^{aa}\,.
\end{equation}
The second term $c^{ba}_{\rm sj}$ encodes inter-band coordinate shift, and can be related to the phonon self energy:
\begin{equation}\label{eq:cbasj}
\small
  c^{ba}_{{\rm sj},y}=(-i) A_{y}^{ba}\left[{\Pi}^{ab}_{-}\left(\ce_a-\frac{i}{2}\Gamma_{a}(k)\right)-{\Pi}^{ab}_{+}\left(\ce_a+\frac{i}{2}\Gamma_a(k)\right)\right]\,.
\end{equation} Here $A^{ba}_\mu$ is the multi-band Berry connection of phonons and ${\Pi}_{\pm}(z)$ is the retarded/advanced phonon self-energy at the complex frequency $z$. Here it is sufficient to calculate $\Pi$ to second order in couplings $K_{ij\alpha}$. We can see from Eq.\eqref{eq:kappaH_semiclassical} and \eqref{eq:cbasj} that shifting $\ce_a\to\ce_a\pm \frac{i}{2}\Gamma_a(k)$ is important to account for the resonance. Otherwise, the phonon self energy would pin $\ce_a=\Delta$ but at this energy the decay rate $\Gamma_{a}(k)$ also diverges.

Finally, we can substitute the phonon dispersion $\ce_{a}=c_{L,T}k$ and evaluate Eq.\eqref{eq:kappaH_semiclassical}, leading to Eq.(8) of the main text.

\subsection{Reference Frame}\label{app:frame}
    In this section we discuss the implications of pyrochlore geometry of $\rm{Pr}_2\rm{Ir}_2\rm{O}_7$ on thermal Hall effect. We assume the phonon decay rate $\Gamma_{a}(k)$ is independent of band index $a$ as in Eq.(9) of the main text. The thermal Hall conductivity depends on the coupling constants $K_{ij\alpha}$ through the combination
\begin{equation}\label{eq:K2exp}
  \kappa_{xy}\propto \frac{K_L}{c_L^3}+\frac{K_T}{c_T^3}\,,
\end{equation} where \cite{Guo2022a}
 \begin{equation}\label{eq:K_L}
    \begin{split}
K_L=  &+2 \left(K_{xy2}+K_{yx2}\right) \left(K_{xx1}-K_{yy1}\right)-2 \left(K_{xy1}+K_{yx1}\right) \left(K_{xx2}-K_{yy2}\right)\\
&+K_{zx1} K_{zy2}-K_{zx2} K_{zy1}\\
&+K_{xz1} K_{yz2}-K_{xz2} K_{yz1}\\
&+K_{zx1} K_{yz2}-K_{zx2} K_{yz1}-K_{xz2} K_{zy1}+K_{xz1} K_{zy2}\,,\\
    \end{split}
    \end{equation}
\begin{equation}\label{eq:K_T}
  \begin{split}
   K_T= &-\frac{5}{2} \left(\left(K_{xy1}+K_{yx1}\right) \left(K_{xx2}-K_{yy2}\right)-\left(K_{xy2}+K_{yx2}\right) \left(K_{xx1}-K_{yy1}\right)\right)\\
&+\frac{1}{2} \left(\left(K_{xy1}-K_{yx1}\right) \left(K_{xx2}+K_{yy2}\right)-\left(K_{xy2}-K_{yx2}\right) \left(K_{xx1}+K_{yy1}\right)\right)\\
&+K_{zx1} K_{zy2}-K_{zx2} K_{zy1}\\
&+4 K_{xz1} K_{yz2}-4 K_{xz2} K_{yz1}\\
&+K_{zz1} \left(K_{xy2}-K_{yx2}\right)+K_{zz2} \left(K_{yx1}-K_{xy1}\right)\,.\\
  \end{split}
\end{equation} The above coupling constants are defined in the global lab frame $H\parallel z$ where the thermal Hall effect is measured. In contrast, the coupling constants from symmetry analysis (Eqs.\eqref{eq:Hspin-ph} and \eqref{eq:Hspin-ph2}) is written in the local frame as defined in Table.~\ref{tab:frame}.

To translate the couplings to the global lab frame, we should perform a rotation on the tensor $K_{ij\alpha}$:
\begin{equation}\label{}
  K_{ij\alpha}^{\text{global},(l)}=\sum_{i',j'}\Lambda^{(l)}_{ii'}\Lambda^{(l)}_{jj'}K_{i'j'\alpha}^{\text{local}}\,.
\end{equation} Here $l=0,1,2,3$ denotes the local frames in Table.~\ref{tab:frame}, and the rotation matrix is
\begin{equation}\label{}
  \Lambda^{(l)}_{ij}=\vec{e}_i\cdot\vec{e}^{(l)}_j\,,
\end{equation} where $\vec{e}_i$ is the basis of the global frame and $\vec{e}^{(l)}_j$ is the basis of the local frame. $K_{ij\alpha}^{\text{local}}$ can be directly read off from Eqs.\eqref{eq:Hspin-ph} and \eqref{eq:Hspin-ph2}:
\begin{equation}\label{}
  K_{ij1}^{\text{local}}=\begin{pmatrix}
            k_1 & 0 & k_{2a} \\
            0 & -k_1 & 0 \\
            k_{2b} & 0 & 0
          \end{pmatrix}\,,
\end{equation}
\begin{equation}\label{}
  K_{ij2}^{\text{local}}=\begin{pmatrix}
                           0 & k_1 & 0 \\
                           k_1 & 0 & -k_{2a} \\
                           0 & -k_{2b} & 0
                         \end{pmatrix}\,.
\end{equation}

Now we can compute $K_L$ and $K_T$ in terms of the irreducible couplings $k_1,k_{2a},k_{2b}$.

When $H\parallel[001]$, we choose $\left(\vec{e}_{i}\right)_{j}=\delta_{ij}$, and we obtain
\begin{equation}\label{eq:KL001}
  K_L^{(0)}=-K_L^{(1)}=-K_L^{(2)}=K_L^{(3)}=\frac{8 k_1^2-(k_{2a}+k_{2b})^2}{\sqrt{3}}\,,
\end{equation}
\begin{equation}\label{}
  K_T^{(0)}=-K_T^{(1)}=-K_T^{(2)}=K_T^{(3)}= \frac{10 k_{1}^2-4k_{2a}^2-k_{2b}^2}{\sqrt{3}}\,.
\end{equation} It may appear that the contribution from the 1st and 2nd sublattice cancel that of the 0th and the 3rd, but there is an additional minus sign from the local $z$-axis $\vec{H}\cdot\vec{z}_0=-\vec{H}\cdot\vec{z}_1=-\vec{H}\cdot\vec{z}_2=\vec{H}\cdot\vec{z}_3=1/\sqrt{3}$, so all four sublattices contribute equally to the thermal Hall effect.

When $H\parallel[111]$, we can choose the global frame to be the zeroth local frame, so we obtain
\begin{equation}\label{}
  K_L^{(0)}=-3K_L^{(1)}=-3K_L^{(2)}=-3K_L^{(3)}=8 k_1^2-(k_{2a}+k_{2b})^2\,,
\end{equation}
\begin{equation}\label{eq:KT111}
  K_T^{(0)}=-3K_T^{(1)}=-3K_T^{(2)}=-3K_T^{(3)}=10 k_{1}^2-4k_{2a}^2-k_{2b}^2\,,
\end{equation} and $\vec{H}\cdot\vec{z}_0=1$, $\vec{H}\cdot\vec{z}_1=\vec{H}\cdot\vec{z}_2=\vec{H}\cdot\vec{z}_3=-1/3$. Therefore the coefficients of 1st,2nd and 3rd sublattices are equal to $1/3$ of that of the 0th.

\subsection{Estimating Hall Angle}\label{app:hallangle}

    We now perform an estimate of the Hall angle. Here we shall ignore the frequency dependence of the phonon decay rate $\Gamma(\omega)$. The error of this approximation can be judged from the temperature dependence of magneto-thermal conductivity, which according to \cite{Uehara2022} is about $20\%$ at 20K. We can now evaluate the  longitudinal thermal conductivity, using Eq. (11) of the main text, yielding
    \begin{equation}\label{}
      \kappa_{xx}=\frac{2\pi^2 k_B^4 T^3}{15 c \Gamma}\,.
    \end{equation}
    Here $c$ is the typical sound velocity and we assume $T$ is much smaller than Debye temperature.

    The phonon thermal Hall effect reads
    \begin{equation}\label{}
      \kappa_{xy}=\frac{1}{30 \pi m}\frac{\Delta^4}{T^2\sinh\left(\Delta/T\right)}\frac{K_2}{\Gamma/c^3}\,,
    \end{equation} where $K_2/c^3$ represents the quadratic combination of coupling constants as given in  Eq.\eqref{eq:K2exp}.

    The Hall angle is therefore (in unit $k_B=1$)
    \begin{equation}\label{}
      \theta=\frac{\kappa_{xy}}{\kappa_{xx}}=\frac{K_2 \Delta^4 }{4\pi^3(mc^2)T^5\sinh\left(\Delta/T\right)}\,.
    \end{equation} Here the ion mass $m=23.8{\rm u}$, implying $mc^2=3.3*10^4{\rm K}$.

    We now evaluate $\theta$ around the maxima of $\kappa_{xx}(T)$ in \cite{Uehara2022}. When $H\parallel[001]$, the maximum of $\kappa_{xx}(T)$ is around 20K, at magnetic field $H=9{\rm T}$. At this magnetic field the Zeeman gap $\Delta=18 {\rm K}$ (We used $gJ=2.6$). We choose $K_2=400^2/3~{\rm K^2}$, and we obtain $\theta=4.1\times10^{-4}$, which agrees with Ref.~\cite{Uehara2022}.

    When the magnetic field is in the $[111]$ direction, the value of $K_2$ is exactly three times larger than that of the $H\parallel[001]$ case (see Eq.\eqref{eq:KL001}-\eqref{eq:KT111}), meaning $K_2=400^2~{\rm K^2}$. The thermal Hall effect is dominated by the pseudospin sitting on the sublattice whose easy axis align with the $[111]$ direction, and at $H=9 {\rm T}$ the Zeeman gap is $\Delta=18\times \sqrt{3} {\rm K}$. Since only one of the four sublattices contribute, the result is multiplied by a factor of $1/4$. At $T=20 {\rm K}$ the Hall angle is $\theta=1.2\times 10^{-3}$. However, the Hall angle is quite sensitive to temperature, and if we instead use $T=23 {\rm K}$, we will obtain $\theta=7.9\times 10^{-4}$, which agrees with the experiment \cite{Uehara2022}.
\section{Suppression of the skew-scattering contribution}\label{app:skew}

In this part we review the arguments that skew scattering is suppressed \cite{Mori2014,Guo2021,Sun2022}.

 The phonon-pseudospin coupling in Eq.(5) of main text has inversion symmetry (under which $\partial_i$ and $u_i$ are flipped), and it couples to phonon modes of odd parity, meaning that the phonon $T$-matrix of scattering on a single pseudospin has the property $T_{k,k'}=-T_{-k,k'}=-T_{k,-k'}$. In our paramagnetic model, there is no correlation between different pseudospins, so the scattering rate $W_{k,k'}$ is the sum of individual $T$-matrix element squared. This implies that $W_{k,k'}=W_{k,-k'}=W_{-k,k'}$. The skew scattering thermal Hall effect of phonons can be computed using Boltzmann equation with an asymmetric component $W^A_{k,k'}=-W^A_{k',k}$ (we assume inversion symmetry) of $W_{k,k'}$, and the result can be written schematically as \cite{Guo2021,Sun2022}
\begin{equation}\label{}
  \kappa_{xy}\sim \int \rd^3k \int \rd^3k' (W^{-1}c_x)_k W^A_{k,k'} (W^{-1}c_y)_{k'}=0\,.
\end{equation} Because the velocity function is of odd parity  $c_{x,k}=-c_{x,-k}$, the $k$ and $k'$ integrals vanish individually.

The above argument can potentially fail if:
\begin{itemize}
  \item The pseudospins are spatially correlated, so that $W_{k,k'}$ contains interference terms of $T$-matrix amplitudes from different pseudospins. This is proposed as a mechanism for thermal Hall effect in terbium garnets \cite{Mori2014}, but within the scope of our model we assume no correlation at leading order of coupling constants. Such correlations might be generated at higher orders of perturbation theory which is probably too small.
  \item There are multiple channels where phonons are coupled to pseudospins and $T_{k,k'}$ might not be equal to $-T_{-k,k'}$ or $-T_{k,-k'}$. This does not happen in our model due to the inversion symmetry of the crystal, which is present for $\rm{Pr}_2\rm{Ir}_2\rm{O}_7$. When the inversion symmetry is present, the coupling must be of the form
      \begin{equation}\label{}
        H'=K_{ij\alpha}\partial_i u^j \sigma^\alpha+K'_{ijkl\alpha}\partial_i\partial_j\partial_k u^l \sigma^\alpha+\dots\,,
      \end{equation} which all contain odd number of derivatives. The $T$-matrix from these couplings satisfies $T_{k,k'}=-T_{-k,k'}=-T_{k,-k'}$.
\end{itemize}

In addition, the above argument does not rely on the fact that $\kappa_{xy}$ is anti-symmetrized, so the same reasoning shows that the vertex correction for the energy current vertex vanishes, and there is no difference between single particle lifetime and transport lifetime at fourth order in $K_{ij\alpha}$.

\section{Details of Fitting}\label{app:fitting}

    We only fit the longitudinal thermal conductivity data $\kappa_{xx}(T,H)$. We minimize the following least square target function ($\gamma$ is the relative weight) $\chi_{H=0}^2+\gamma\chi_\text{MTC}^2$, where
    \begin{equation}\label{}
      \chi_{H=0}^2=\sum_{i} \left(\frac{\kappa_{xx,\text{data},i}(T_i,H=0)-\kappa_{xx,\text{model},i}(T_i,H=0)}{\kappa_{xx,\text{data},i}(T_i,H=0)}\right)^2\,,
    \end{equation}  and
    \begin{equation}\label{}
      \chi_{\text{MTC}}^2=\sum_{i} w_i\left(\frac{\kappa_{xx,\text{model},i}(T_i,H_i)}{\kappa_{xx,\text{model},i}(T_i,H=0)}-\frac{\kappa_{xx,\text{data},i}(T_i,H_i)}{\kappa_{xx,\text{data},i}(T_i,H=0)}\right)^2\,.
    \end{equation} Here $i$ indexes data points. Because we don't have the uncertainty of the zero field $\kappa_{xx}$, we assume a poisson fluctuation of the data to obtain a dimensionless $\chi_{H=0}^2$. Within $\chi_{\text{MTC}}^2$ we weight data at low field ($\kappa_{xx}(H)<\kappa_{xx}(0)$) with weight $w_i=1$ and data at high field ($\kappa_{xx}(H)>\kappa_{xx}(0)$) with weight $w_i=0.15$.

    For the case of $H\parallel[001]$, we set $\gamma=1$ and obtained reasonable agreement with both $\kappa_{xx}(H=0)$ and MTC data. The scattering rate $\Gamma(\omega)$ is in unit $\rs^{-1}$ and frequency $\omega$ is in unit $\rK$. We obtain $R=2.2\times 10^7 \rs^{-1}\rK^{-2}$, $A_0=6.4\times 10^8 \rs^{-1}\rK^{-1}$, $A_1=3.0\times 10^{-9}\rs^{-1}\rK^{-4}$, $A_2=1.4\times 10^6 \rs^{-1}\rK^{-3}$ and $gJ=3.42$.  The non-resonant scattering is dominant by linear defect scattering ($A_0$), mainly because the low-$T$ $\kappa_{xx}(T,H=0)$ is approximately quadratic in $T$.

    For the case of $H\parallel[111]$, we can not simultaneously fit $\kappa_{xx}(H=0)$ and MTC, and we choose $\gamma=50$ to focus on the MTC data. We obtain $R_2=3R_1=1.5\times 10^7 \rs^{-1}\rK^{-2}$, $A_0=8.9\times 10^8 \rs^{-1}\rK^{-1}$, $A_1\approx 0$, $A_2=4.6\times 10^5 \rs^{-1}\rK^{-3}$ and $gJ=2.38$.

    We compare our extracted $R$ value to Eq.(7) of the main text. We take the ion mass $m=23.8{\rm u}$, which is the harmonic average of $\rm{Pr}_2\rm{Ir}_2\rm{O}_7$. The speed of sound is $c=3424{\rm m/s}$ and the Debye frequency is $\omega_D=436\rK$. We assume the coupling constant $K=800\rK$ as estimated in previous section. Then we have (keep in mind $\hbar=k_B=1$)
    \begin{equation}\label{}
      R_\text{theory}=A_R\frac{K^4}{(mc^2)^2 \omega_D^3}=5.7\times 10^5 A_R \rs^{-1}\rK^{-2}\,.
    \end{equation} Here $A_R$ is a numerical pre-factor. To match with the fitting we need $A_R\approx 40$, which suggests that the fitting result overestimates it.
\end{widetext}
\bibliography{phonon}
\end{document}